\batchmode

\documentclass[12pt,preprint]{aastex}

\newcommand{\be}{\begin{equation}}
\newcommand{\ee}{\end{equation}}
\newcommand{\ul}{\underline{\hspace{40pt}}}
\newcommand{\vs}{\mbox{$v_{\rm{s}}$}}
\newcommand{\vims}{\mbox{$v_{\rm ims}$}}

\newcommand{\vms}{\mbox{$v_{\rm ms}$}}
\newcommand{\Mna}{\mbox{$M_{\rm nA}$}}
\newcommand{\vdot}{\mbox{\boldmath{$\cdot$}}}
\newcommand{\cross}{\mbox{\boldmath{$\times$}}}
\newcommand{\grad}{\mbox{\boldmath{$\nabla$}}}
\newcommand{\curl}{\mbox{\grad\cross}}
\newcommand{\xhat}{\mbox{\boldmath{$\hat{x}$}}}
\newcommand{\jvec}{\mbox{\boldmath{$j$}}}

\newcommand{\Evec}{\mbox{\boldmath{$E$}}}
\newcommand{\Fvec}{\mbox{\boldmath{${\cal F}$}}}
\newcommand{\yvec}{\mbox{\boldmath{$y$}}}
\newcommand{\fvec}{\mbox{\boldmath{$f$}}}
\newcommand{\lvec}{\mbox{\boldmath{$l$}}}
\newcommand{\Cvec}{\mbox{\boldmath{${\cal C}$}}}
\newcommand{\Cvecb}{\mbox{\boldmath{${\cal C}^{\prime}$}}}
\newcommand{\Uvec}{\mbox{\boldmath{${\cal U}$}}}
\newcommand{\Uvecb}{\mbox{\boldmath{${\cal U}^{\prime}$}}}
\newcommand{\Amatrix}{\mbox{\boldmath{${\sf A}$}}}
\newcommand{\simlt}{\lesssim}
\newcommand{\simgt}{\gtrsim}
\newcommand{\Order}{\mbox{${\cal O}$}}
\newcommand{\dz}{\mbox{$\Delta z$}}
\newcommand{\zbar}{\mbox{$\overline{z}$}}
\newcommand{\fbar}{\mbox{$\overline{f}$}}
\newcommand{\vbar}{\mbox{$\overline{v}$}}
\newcommand{\delfbar}{\mbox{$(\overline{\Delta f})$}}
\newcommand{\fR}{\mbox{$f_{{}_{R}}$}}
\newcommand{\vR}{\mbox{$v_{{}_{R}}$}}
\newcommand{\fL}{\mbox{$f_{{}_{L}}$}}
\newcommand{\vL}{\mbox{$v_{{}_{L}}$}}
\newcommand{\cdotm}{\mbox{\boldmath{$\cdot$}}}
\newcommand{\Qvisc}{\mbox{$Q_{\rm{visc}}$}}
\newcommand{\kB}{\mbox{$k_{\rm B}$}}
\newcommand{\cmMMM}{\mbox{cm$^{-3}$}}
\newcommand{\gcmMMM}{\mbox{g\,cm$^{-3}$}}
\newcommand{\cc}{\mbox{cm$^{-3}$}}
\newcommand{\kms}{\mbox{km s$^{-1}$}}
\newcommand{\muf}{\mbox{$m_{\rm f}$}}
\newcommand{\nf}{\mbox{$n_{\rm f}$}}
\newcommand{\rhof}{\mbox{$\rho_{\rm f}$}}
\newcommand{\vfvec}{\mbox{\boldmath$v$}\mbox{$_{\rm f}$}}
\newcommand{\Omegafvec}{\mbox{\boldmath$\Omega$}\,\mbox{$_{\rm f}$}}
\newcommand{\Ef}{\mbox{$E_{\rm f}$}}
\newcommand{\Pf}{\mbox{$P_{\rm f}$}}

\newcommand{\tdij}{\mbox{$\tau^{\rm d}_{\rm i,j}$}}
\newcommand{\tcfij}{\mbox{$\tau^{\rm cf}_{\rm i,j}$}}
\newcommand{\vshk}{\mbox{$v_{\rm{s}}$}}
\newcommand{\mun}{\mbox{$m_{\rm n}$}}
\newcommand{\nn}{\mbox{$n_{\rm n}$}}
\newcommand{\nno}{\mbox{$n_{\rm n0}$}}
\newcommand{\rhon}{\mbox{$\rho_{\rm n}$}}
\newcommand{\rhono}{\mbox{$\rho_{\rm n0}$}}
\newcommand{\Tn}{\mbox{$T_{\rm n}$}}
\newcommand{\Tno}{\mbox{$T_{\rm{n0}}$}}
\newcommand{\Pn}{\mbox{$P_{\rm n}$}}
\newcommand{\En}{\mbox{$E_{\rm n}$}}
\newcommand{\vn}{\mbox{$v_{{\rm n}}$}}
\newcommand{\vnvec}{\mbox{\boldmath$v$} \mbox{$_{\rm n}$}}
\newcommand{\vny}{\mbox{$v_{{\rm n}y}$}}
\newcommand{\vnz}{\mbox{$v_{{\rm n}z}$}}
\newcommand{\Sn}{\mbox{$S_{\rm n}$}}
\newcommand{\Gn}{\mbox{$G_{\rm n}$}}
\newcommand{\Ln}{\mbox{$\Lambda_{\rm n}$}}
\newcommand{\vecFn}{\mbox{\boldmath$F$}\mbox{$_{\rm n}$}}
\newcommand{\Fny}{\mbox{$F_{{\rm n}y}$}}
\newcommand{\Fnz}{\mbox{$F_{{\rm n}z}$}}
\newcommand{\Lneut}{\mbox{$L_{{\rm neut}}$}}
\newcommand{\Czero}{\mbox{$C_{\rm{n0}}$}}
\newcommand{\machno}{\mbox{$M$}}
\newcommand{\vano}{\mbox{$v_{\rm{A,n0}}$}}
\newcommand{\vnzsf}{\mbox{$v_{{\rm n}z,{\rm{shk}}}$}}
\newcommand{\Bvec}{\mbox{\boldmath{$B$}}}
\newcommand{\Bx}{\mbox{$B_x$}}
\newcommand{\Lfield}{\mbox{$L_{{\rm field}}$}}
\newcommand{\Bo}{\mbox{$B_{0}$}}
\newcommand{\me}{\mbox{$m_{\rm e}$}}
\newcommand{\nne}{\mbox{$n_{\rm e}$}}
\newcommand{\rhoe}{\mbox{$\rho_{\rm e}$}}

\newcommand{\xxeo}{\mbox{$x_{\rm{e0}}$}}
\newcommand{\vevec}{\mbox{\boldmath$v$}\mbox{$_{\rm e}$}}
\newcommand{\tdragen}{\mbox{$\tau^{\rm d}_{\rm e,n}$}}

\newcommand{\Ge}{\mbox{$G_{\rm e}$}}
\newcommand{\Le}{\mbox{$\Lambda_{\rm e}$}}
\newcommand{\Te}{\mbox{$T_{\rm e}$}}
\newcommand{\mi}{\mbox{$m_{\rm i}$}}
\newcommand{\nni}{\mbox{$n_{\rm i}$}}
\newcommand{\Ti}{\mbox{$T_{\rm i}$}}

\newcommand{\rhoi}{\mbox{$\rho_{\rm i}$}}
\newcommand{\vivec}{\mbox{\boldmath$v$}\mbox{$_{\rm i}$}}
\newcommand{\tdragin}{\mbox{$\tau^{\rm d}_{\rm i,n}$}}
\newcommand{\vi}{\mbox{$v_{{\rm i}}$}}
\newcommand{\viy}{\mbox{$v_{{\rm i}y}$}}
\newcommand{\viz}{\mbox{$v_{{\rm i}z}$}}
\newcommand{\Halli}{\mbox{$\Gamma_{\rm i}$}}

\newcommand{\sigvin}{\mbox{$\left<\sigma v\right>_{\rm in}$}}
\newcommand{\siggeo}{\mbox{$\sigma_{\rm{en}}$}}
\newcommand{\aplus}{\mbox{a$^+$}}
\newcommand{\maplus}{\mbox{$m_{{\rm a}^+}$}}
\newcommand{\mplus}{\mbox{m$^+$}}
\newcommand{\mmplus}{\mbox{$m_{{\rm m}^+}$}}

\newcommand{\xxio}{\mbox{$x_{\rm{i0}}$}}
\newcommand{\vizsf}{\mbox{$v_{{\rm i}z,{\rm{shk}}}$}}
\newcommand{\rhosg}{\mbox{$\rho_{\rm sg}$}}
\newcommand{\asg}{\mbox{$a_{\rm sg}$}}

\newcommand{\xsg}{\mbox{$x_{\rm sg}$}}
\newcommand{\xsgo}{\mbox{$x_{\rm {sg,0}}$}}
\newcommand{\Ldragsg}{\mbox{$L_{\rm drag}({\rm sg})$}}

\newcommand{\chisgo}{\mbox{$\chi_{\rm{sg,0}}$}}
\newcommand{\Omegasg}{\mbox{$\Omega_{\rm{sg}}$}}
\newcommand{\sgminus}{\mbox{sg$^{-}$}}
\newcommand{\Ssgm}{\mbox{$S_{{\rm sg}^-}$}}
\newcommand{\nsgm}{\mbox{$n_{{\rm sg}^-}$}}
\newcommand{\rhosgm}{\mbox{$\rho_{{\rm sg}^-}$}}
\newcommand{\vsgmvec}{\mbox{\boldmath$v$} \mbox{$_{{\rm sg}^-}$}}
\newcommand{\vsgmz}{\mbox{$v_{{\rm sg}^{-}z}$}}
\newcommand{\Omegasgmvec}{\mbox{\boldmath$\Omega$} \mbox{$_{{\rm sg}^-}$}}
\newcommand{\tdragsgmn}{\mbox{$\tau^{\rm d}_{{\rm sg}^-,{\rm n} }$}}
\newcommand{\tcfsgmi}{\mbox{$\tau^{\rm cf}_{{\rm sg}^-,{\rm i} }$}}

\newcommand{\xsgmo}{\mbox{$x_{{\rm sg}^-,0}$}}
\newcommand{\sgzero}{\mbox{sg$^{0}$}}
\newcommand{\Ssgz}{\mbox{$S_{{\rm sg}^0}$}}

\newcommand{\rhosgz}{\mbox{$\rho_{{\rm sg}^0}$}}
\newcommand{\vsgzvec}{\mbox{\boldmath$v$} \mbox{$_{{\rm sg}^0}$}}
\newcommand{\vsgzz}{\mbox{$v_{{\rm sg}^{0}z}$}}
\newcommand{\tdragsgzn}{\mbox{$\tau^{\rm d}_{{\rm sg}^0,{\rm n} }$}}
\newcommand{\tcfsgze}{\mbox{$\tau^{\rm cf}_{{\rm sg}^0,{\rm e} }$}}
\newcommand{\tcfsgzi}{\mbox{$\tau^{\rm cf}_{{\rm sg}^0,{\rm i} }$}}

\newcommand{\xsgzo}{\mbox{$x_{{\rm sg}^0,0}$}}
\newcommand{\sgplus}{\mbox{sg$^{+}$}}
\newcommand{\Ssgp}{\mbox{$S_{{\rm sg}^+}$}}
\newcommand{\nsgp}{\mbox{$n_{{\rm sg}^+}$}}
\newcommand{\rhosgp}{\mbox{$\rho_{{\rm sg}^+}$}}
\newcommand{\vsgpvec}{\mbox{\boldmath$v$} \mbox{$_{{\rm sg}^+}$}}
\newcommand{\vsgpz}{\mbox{$v_{{\rm sg}^{+}z}$}}
\newcommand{\Omegasgpvec}{\mbox{\boldmath$\Omega$} \mbox{$_{{\rm sg}^+}$}}
\newcommand{\tdragsgpn}{\mbox{$\tau^{\rm d}_{{\rm sg}^+,{\rm n} }$}}
\newcommand{\tcfsgpe}{\mbox{$\tau^{\rm cf}_{{\rm sg}^+,{\rm e} }$}}

\newcommand{\xsgpo}{\mbox{$x_{{\rm sg}^+,0}$}}
\newcommand{\mg}{m_{\rm g}}
\newcommand{\rhog}{\mbox{$\rho_{\rm g}$}}
\newcommand{\gminus}{\mbox{g$^{-}$}}
\newcommand{\gzero}{\mbox{g$^{0}$}}
\newcommand{\gplus}{\mbox{g$^{+}$}}
\newcommand{\ag}{\mbox{$a_{\rm g}$}}

\newcommand{\xg}{\mbox{$x_{\rm g}$}}
\newcommand{\xgo}{\mbox{$x_{\rm{g,0}}$}}

\newcommand{\chigo}{\mbox{$\chi_{\rm{g,0}}$}}
\newcommand{\Ldragg}{\mbox{$L_{\rm drag}({\rm g})$}}
\newcommand{\Omegag}{\mbox{$\Omega_{\rm g}$}}
\newcommand{\ngm}{\mbox{$n_{{\rm g}^-}$}}
\newcommand{\rhogm}{\mbox{$\rho_{{\rm g}^-}$}}
\newcommand{\vgmvec}{\mbox{\boldmath$v$} \mbox{$_{{\rm g}^-}$}}
\newcommand{\vgmy}{\mbox{$v_{{\rm g}^{-}y}$}}
\newcommand{\vgmz}{\mbox{$v_{{\rm g}^{-}z}$}}
\newcommand{\Sgm}{\mbox{$S_{{\rm g}^-}$}}

\newcommand{\tdraggmn}{\mbox{$\tau^{\rm d}_{{\rm g}^-,{\rm n} }$}}
\newcommand{\tcfgmi}{\mbox{$\tau^{\rm cf}_{{\rm g}^-,{\rm i} }$}}

\newcommand{\xgmo}{\mbox{$x_{{\rm g}^-,0}$}}
\newcommand{\rhogz}{\mbox{$\rho_{{\rm g}^0}$}}

\newcommand{\vgzy}{\mbox{$v_{{\rm g}^{0}y}$}}
\newcommand{\vgzz}{\mbox{$v_{{\rm g}^{0}z}$}}
\newcommand{\Sgz}{\mbox{$S_{{\rm g}^0}$}}

\newcommand{\tdraggzn}{\mbox{$\tau^{\rm d}_{{\rm g}^0,{\rm n} }$}}
\newcommand{\tcfgze}{\mbox{$\tau^{\rm cf}_{{\rm g}^0,{\rm e} }$}}
\newcommand{\tcfgzi}{\mbox{$\tau^{\rm cf}_{{\rm g}^0,{\rm i} }$}}

\newcommand{\xgzo}{\mbox{$x_{{\rm g}^0,0}$}}
\newcommand{\ngp}{\mbox{$n_{{\rm g}^+}$}}
\newcommand{\rhogp}{\mbox{$\rho_{{\rm g}^+}$}}
\newcommand{\vgpvec}{\mbox{\boldmath$v$} \mbox{$_{{\rm g}^+}$}}
\newcommand{\vgpy}{\mbox{$v_{{\rm g}^{+}y}$}}
\newcommand{\vgpz}{\mbox{$v_{{\rm g}^{+}z}$}}
\newcommand{\Sgp}{\mbox{$S_{{\rm g}^+}$}}

\newcommand{\tdraggpn}{\mbox{$\tau^{\rm d}_{{\rm g}^+,{\rm n} }$}}
\newcommand{\tcfgpe}{\mbox{$\tau^{\rm cf}_{{\rm g}^+,{\rm e} }$}}

\newcommand{\xgpo}{\mbox{$x_{{\rm g}^+,0}$}}
\newcommand{\xmetal}{\mbox{$x_{\rm a}$}}
\newcommand{\nH}{\mbox{$n$}}
\newcommand{\water}{\mbox{H$_2$O}}
\newcommand{\Htwo}{\mbox{H$_2$}}

\newcommand{\tbnt}{\tablenotetext}
\newcommand{\photon}{\mbox{h$\nu$}}
\newcommand{\plus}{\mbox{~~$+$~~}}
\newcommand{\goesto}{\mbox{~~$\longrightarrow$~~}}

\shorttitle{MHD Shocks with Grain Dynamics}
\shortauthors{G. E. Ciolek and W. G. Roberge}

\begin{document}

\title{Time-Dependent, Multifluid, Magnetohydrodynamic Shock Waves with
Grain Dynamics I. Formulation and Numerical Tests}

\author{Glenn E. Ciolek and Wayne G. Roberge}

\affil{New York Center for Studies on the Origin of Life (NSCORT) \\
and \\
Department of Physics, Applied Physics and Astronomy,
Rensselaer Polytechnic Institute,  \\ 110 8th Street, Troy, NY 12180}
\email{cioleg@rpi.edu, roberw@rpi.edu}

\begin{abstract}
This is the first in a series of papers on the effects of dust
on the formation, propagation, and structure of nonlinear MHD waves and
MHD shocks in weakly-ionized plasmas.
We model the plasma as a system of 9 interacting fluids, consisting of
the neutral gas, ions, electrons, and 6 grain fluids
comprised of very small grains or PAHs and classical
grains in different charge states.
We formulate the governing equations for
perpendicular shocks under approximations appropriate for dense molecular
clouds, protostellar cores, and protoplanetary disks.
We describe a code that obtains numerical solutions
using a finite difference method, and establish its accuracy by
comparing numerical and exact solutions for special cases.
\end{abstract}



\section{Introduction}

Shock waves in weakly-ionized plasmas usually exhibit a multifluid
structure, in which the charged and neutral particles behave as
separate, interpenetrating fluids. If the shock speed is less than the
speed at which the charged fluid communicates compressive disturbances, 
the charged particles will be accelerated ahead of the neutrals in
a ``magnetic precursor,'' and the neutrals accelerated,
compressed and heated downstream by collisions with streaming
charged particles (Mullan 1971; Draine 1980, henceforth D80).
The flow is continuous (a C- or C$^*$ shock) if the
length scale for acceleration of the neutrals is large compared to
the scale for cooling; otherwise the flow undergoes a discontinuous
transition (a J shock) at a viscous subshock
(D80; Chernoff 1987; Roberge \& Draine 1990).
The dynamics, chemistry, and emission from multifluid
shock waves can differ profoundly from those of their single-fluid
counterparts. Consequently, they have been the subject of numerous
investigations (see Draine \& McKee 1993 for the
most recent comprehensive review).

The energy and momentum of the ``charged fluid'' are usually
dominated by the magnetic field, the charged particles acting
mainly to provide energy and momentum transfer between the field
and the neutral fluid.
At low densities the coupling is dominated by ion-neutral
scattering and shock waves can be modeled as
a three-fluid system: the neutral gas, the ion/magnetic field
fluid, and the fluid of electrons (which move with the ions
but need not have the same temperature; see D80).
However when the density of the preshock medium exceeds
$\nH \sim 10^5$\,\cmMMM,\footnote{We use \nH\ to denote the
density of H nuclei in all forms.} the energy and momentum
transfer is dominated by collisions between neutral particles and
charged dust grains (Elmegreen 1979; D80; Nakano \& Umebayashi 1980;
Ciolek \& Mouschovias 1993).
If the grains are well coupled to the magnetic field, they
can be incorporated into the 3-fluid picture as an additional
species of ion.
This is a reasonable approximation for PAHs and very small grains
with radii $\asg \sim 10^{-7}$\,cm
(henceforth referred to collectively as ``small grains'') at
densities up to $\sim 10^9$ (Neufeld \& Hollenbach 1994).
If the small grains are present in dense gas with a fractional
abundance not much less than $\sim 10^{-7}$
(the abundance inferred for the diffuse ISM, see Li \& Draine 2001 and
references therein), they will dominate the field-neutral coupling and
a 3-fluid model of the dynamics is appropriate
(e.g., Kaufman \& Neufeld 1996a,b).

However if small grains are depleted in dense gas, as seems likely, then
scattering by large grains with $\ag \simgt 0.1$\,\micron\ 
will dominate the field-neutral coupling.
Because the large grains are only partially tied to the magnetic
field for $\nH \simgt 10^5$\,\cmMMM, they must be treated as an
additional fluid (Chernoff 1985; Havnes, Hartquist \& Pilipp 1987;
Ciolek \& Mouschovias 1993).
The problem is further complicated by the fact that large grains
have different charge states, a continuous spectrum of sizes, 
and masses many orders of magnitude larger than those of the ions.
Draine (D80) developed an approximate description of the grain dynamics
by treating the grains as test particles and neglecting their
inertia; in this approximation each grain drifts through the
neutrals with a speed that can be determined by balancing the gas drag
and electromagnetic forces. Chernoff (1985)
treated the grains as a separate fluid, included their inertia, and
demonstrated that the grains can undergo a J shock.
Pilipp, Hartquist \& Havnes (1990) and Pilipp \& Hartquist (1994)
carried out numerical simulations of steady 4-fluid shocks,
emphasizing the importance of grain charging processes
and the dynamical effects of the grain charge and current.
Wardle (1998, see also Wardle \& Ng 1999) has formulated 
equations of motion valid for steady multifluid shocks with
an arbitrary number
of grain fluids, in a form that is valid for very large
densities, and has discussed optimal procedures for representing
the spectrum of large grain sizes by a single, effective size.

This is the first in a series of papers on the effects of dust
on the formation, propagation, and structure of nonlinear MHD waves
and shocks in dense molecular clouds, bipolar outflows, and
protoplanetary disks. Here we describe a dynamical model that
incorporates and extends previous work in several ways:
(i) We allow for time-dependent flow. Prior work on
time-dependent, multifluid shocks (T\'{o}th 1994; Smith \& Mac Low 1997;
Mac Low \& Smith 1997; Stone 1997;
Chieze, Pineau des Forets \& Flower 1998) has been based on the
3-fluid model. We are interested in time-dependent solutions
because they permit us to study the formation of magnetic precursors,
because of their possible relevance to bipolar outflows
(Flower \& Pineau des Forets 1999),
and because time-dependent simulations are a robust way
to find steady solutions (e.g., steady J shocks) that are otherwise
hard to compute.
(ii) We include the inertia of large grains, which has
been neglected in all prior studies except Chernoff's (1985).
Neglecting the inertia is a good approximation when the
gas drag time for large grains is small compared
to all other dynamical time scales.
While this is usually appropriate for the dynamical time of
the neutrals, it is often
a poor approximation for the charged fluid (see \S2.3.3).
A realistic treatment of grain dynamics is necessary
for studies on the formation of magnetic precursors:
Draine \& McKee (1993) pointed out that magnetic precursors
could not form (and C shocks therefore could not exist) 
for shocks faster than $\approx 20~\kms$ if 
large grains are able to ``load'' the magnetic field lines efficiently.
The code we describe was designed in part to address this issue.
(iii) We include small grains as a separate fluid.
This permits us to model grain charging realistically
and to examine various scenarios for small-grain
depletion in dense gas.
(iv) We include charge fluctuations by treating the grains
in different charge states (3 each for small- and large grains)
as separate fluids. Charge fluctuations exchange mass,
momentum, and energy between the charged and neutral grain
fluids, with implications for the loading of magnetic field lines.
Nevertheless, the scope of our calculations is limited in
some important respects:
We do not attempt to describe the continuous size distribution of
the large grains, which are assumed to have uniform radii.
We have also restricted our initial investigations to one-dimensional
geometry so that, for example, time-dependent simulations of Wardle
instabilities with dust are beyond our present capabilities.

The plan of this paper is as follows. In \S~2 we give the equations of
motion and discuss the conditions where various approximations are
likely to be valid. In \S~3 we describe our numerical methods and
summarize a suite of benchmark calculations that establish the accuracy
of our code. Our results are summarized in \S~4.


\section{Formulation}

\subsection{Modeling Assumptions}

We consider plane-parallel disturbances propagating along
the $+z$ direction of a
Cartesian frame, $\left(x,y,z\right)$, defined so that the undisturbed
plasma at $z=+\infty$ is at rest.
Fluid variables are assumed to depend only on $z$ and $t$.
The plasma is threaded by a magnetic field of the form
$\Bvec = \Bx(z,t) \xhat$.
That is we restrict the discussion to perpendicular shocks;
oblique shocks will be described elsewhere.
Fluid motions in the $x$-direction (i.e., along the field) are
uncoupled from those in the $y$ and $z$ directions and are
ignored.

We model the plasma as a system of 9 interacting fluids.
Fluid variables are labeled by a subscript f, where
${\rm f} = {\rm n}$, i, e, \sgminus, \sgzero, \sgplus, \gminus,
\gzero, and \gplus\ denote the neutral fluid, ions, electrons,
small grains with charge $-e$, $0$, or $+e$, and large
grains with charge $-e$, $0$, or $+e$, respectively.
The particles of fluid f have mean mass \muf,
number density \nf, mass density \rhof,
flow velocity \vfvec, pressure \Pf, and internal energy density \Ef,
where \Ef\ includes the contributions of thermal
motions, rotational and vibrational excitation, dissociation, etc.

The neutral and ion fluids are composed of $N_{\rm n}$ and
$N_{\rm i}$ distinct chemical species, respectively, so that
\be
\nn = \sum_{\alpha=1}^{N_{\rm n}} \,n_{n\alpha}
\ee
and similarly for the ions.
Realistic simulations require chemical networks elaborate enough
to predict the abundances of important coolants
(primarily \water, CO, and \Htwo\ in dense plasmas;
see Neufeld \& Kaufman 1993; Neufeld, Lepp \& Melnick 1995) and to model
changes in the fractional ionization (which can profoundly affect the
hydrodynamics; see Flower, Pineau des For\^{e}ts \& Hartquist 1985).
In subsequent papers we include chemical networks tailored to the
particular applications of interest to us.
Here we are interested mainly in establishing the accuracy of our 
numerical code.
The benchmark calculations we present were obtained using
a simple analytic expression for the radiative cooling rate (see \S3);
the chemistry of coolants is therefore not required.
To model the ionization we adopt a drastically simplified ``chemistry'' 
with just 4 neutral species and 2 ions.
The neutral fluid is composed of molecular hydrogen
with $n_{{\rm H}_2}=0.5\nH$, atomic helium with $n_{\rm He}=0.1\nH$,
plus a generic heavy atom, a, and molecule, m.
The ion fluid is composed of the atomic ion \aplus\ and
molecular ion \mplus.
Since the dominant ions in the dense astrophysical plasmas relevant
to star formation and protostellar disks are HCO$^+$, Na$^+$, and Mg$^+$,
we take the ion mass to be $\maplus=\mmplus=\mi = 25$\,amu.
The chemical abundances are determined by a network
of 20 gas-phase and grain surface reactions (Appendix~A). 
The fractional ionization predicted by our simple network is
in good agreement with more elaborate models
for $\nH \simgt 10^5$\,\cmMMM\ (Fig.~1).

The large and small grains are assumed to have uniform radii
\ag\ and \asg, where $\ag \sim 10^{-5}$\,cm and $\asg\sim 10^{-7}$\,cm
are adjustable parameters. Modifying our code to include a range of
grain sizes would be straightforward but is beyond our current
computational resources. The total abundances of large and small grains,
summed over all charge states, are assumed constant.

\subsection{Governing Equations}

\subsubsection{Neutral Particles}

The mass, momentum, and energy equations for the neutral fluid are
\be
\frac{\partial\rho_{\rm n}}{\partial t}
+
\frac{\partial}{\partial z}
\left( \rhon v_{{\rm n}z} \right)
=
\Sn,
\label{n_mass_eq}
\ee
\be
\frac{\partial}{\partial t}
\left( \rhon\vnz \right)
+
\frac{\partial}{\partial z}
\left( \rhon v_{\rm n}^2+\Pn+\Qvisc \right)
=
\Fnz,
\label{n_zmtm_eq}
\ee
\be
\frac{\partial}{\partial t}
\left( \rhon\vny \right) 
+
\frac{\partial}{\partial z}
\left( \rhon\vny\vnz \right)
=
\Fny,
\label{n_ymtm_eq}
\ee
and
\be
\frac{\partial}{\partial t}
\left( \frac{1}{2}\rhon v_{\rm n}^2+\En \right)
+
\frac{\partial}{\partial z}
\left[\,
\left( \frac{1}{2}\rhon v_{\rm n}^2+\En+\Pn+\Qvisc \right)\vnz
\,\right]
=
\vecFn\vdot\vnvec + \Gn - \Ln
\label{n_energy_eq}
\ee
respectively, where $\vn^{2} = \vnz^{2} + \vny^{2}$; the
``source terms'' \Sn, \vecFn\ and \Gn\ are the net rates per volume at
which mass, momentum and thermal energy are added to the neutrals by
interactions with the other fluids, and \Ln\ is the rate of radiative
cooling.
%
%
%
%
The benchmark tests presented in \S3 were carried out for
an idealized fluid with \Ln\ prescribed analytically and
$\En = \Pn/\left(\gamma-1\right)$, with constant $\gamma$.
Future papers will incorporate a realistic calculation of radiative
cooling by CO, $\water$, and $\Htwo$ (Neufeld \& Kaufman 1993;
Neufeld, Lepp \& Melnick 1995) and the internal energy of $\Htwo$
accounting for its time-dependent excitation and $\Htwo$ dissociation.
Our model includes artificial
viscosity to smear out jump fronts in the neutral flow. We use the
von Neumann-Richtmyer prescription for the pseudo-viscous stress,
\be
\Qvisc = \left\{
\begin{array}{cl}
q \left(\dz \right)^2\,\rho_{\rm n}\,
\left|\partial v_n/\partial z\right|^2 &
      {\rm if}~\partial v_n/\partial z<0,    \\
 & \\
0  & {\rm otherwise} \\
\end{array}
\right.
\label{Qvisc_eq}
\ee
(see Richtmyer \& Morton 1967),
where \dz\ is the grid spacing and $q$ is an adjustable
dimensionless parameter. If present, a jump front will be smeared
over $\sim q$ grid cells.

\subsubsection{Electrodynamics}

We assume that the magnetic field is frozen into the fluid
of ions and electrons.
(The validity of this approximation is discussed in \S2.3.)
In the frozen field approximation, the electrons and ions
move with a common velocity,
\be
\vevec = \vivec,
\label{ve_eq}
\ee
the electric field satisfies
\be
\Evec \approx -{\vivec\over c} \cross \Bvec,
\label{Efield_eq}
\ee
and the  Maxwell equations reduce to the induction equation,
\be
\frac{\partial B_x}{\partial t}
+
\frac{\partial}{\partial z}\left(B_x v_{{\rm i}z}\right)
=
0,
\label{induction_eq}
\ee
plus the requirement of macroscopic charge neutrality,
\be
\nne = \nni+\ngp+\nsgp-\ngm-\nsgm.
\label{neutrality_eq}
\ee

\subsubsection{Electrons}

We do not solve mass or momentum equations for the electrons
because \nne\ and \vevec\ are functions of the other
fluid variables in the frozen-field approximation
(cf.\ eqs.\ [\ref{ve_eq}] and [\ref{neutrality_eq}]).
The electron energy equation is omitted in favor of Chernoff's
(1987) approximation,
\be
\Ge=\Le,
\label{Te_eq}
\ee
which says that the electron fluid rapidly reaches thermal balance
due to its small heat capacity.
%
The shock structure depends on the electron temperature via
the rate coefficients for electron impact excitation of molecules
(which contribute to the radiative cooling) and the rate
coefficients for recombination and grain charging (which influence
the ionization; see Appendix~A).
The solutions presented in \S3 are insensitive to $\Te$ because
the cooling is prescribed analytically and the flow time is much
less than the timescale for changes in the ionization.
We therefore adopt the (unrealistic) approximation $\Te=0.15 \Ti$
for expediency; this approximation is based on the results of Draine,
Roberge, \& Dalgarno (1983, hereafter DRD83). Future papers on
realistic simulations will incorporate a detailed description of
electron heating and cooling including the atomic and molecular
processes described in DRD83.

\subsubsection{Ions}

We solve the mass equation for the ions,
\be
\frac{\partial \rhoi}{\partial t}
+
\frac{\partial}{\partial z}
\left( \rhoi \viz \right)
=
-\Sn,
\label{i_mass_eq}
\ee
but not the momentum or energy equations.
We calculate the momentum of the ion fluid by (i) neglecting the
inertia of the ions and electrons and (ii) requiring the total
electric current to be consistent with the magnetic field gradient.
We assume that the ion and electron velocities are determined
solely by the balance between electromagnetic forces and drag
due to elastic collisions with the neutrals.
Then requirement (i) implies
\be
0 = -e\nne\left(\Evec+{\vevec\over c}\cross\Bvec\right)
    +\frac{\rhoe}{\tdragen}\left(\vnvec-\vevec\right)
\label{vedrift_eq}
\ee
and
\be
0 = e\nni\left(\Evec+{\vivec\over c}\cross\Bvec\right)
    +\frac{\rhoi}{\tdragin}\left(\vnvec-\vivec\right),
\label{vidrift_eq}
\ee
where \tdij\ is the drag time for species i due to
elastic scattering with species j (a function of the
fluid variables, see DRD83).\footnote{Notice that
equations (\ref{vedrift_eq}) and
(\ref{vidrift_eq}) would yield the nonsensical solution
$\vevec=\vivec=\vnvec$ if one naively inserted expression
(\ref{Efield_eq}) for the electric field.
The meaning of this result is that the drag force must
vanish in the frozen-field approximation, where gas drag
is taken to be negligible compared with electromagnetic
forces.}
Adding expressions (\ref{vedrift_eq}) and (\ref{vidrift_eq})
gives
\be
e\left(\nni-\nne\right)\Evec
+{e\over c}\left(\nni\vivec-\nne\vevec\right)\cross\Bvec
+\frac{\rhoi}{\tdragin}\left(\vnvec-\vivec\right)
\approx 0,
\label{sumdrft_eq}
\ee
where we have neglected the drag force on the electron fluid
compared to the drag force on the ions.\footnote{The ratio
of the drag forces is of order
$(\me/\mun)\siggeo\bar{v}_{\rm e}/\sigvin \simlt 10^{-3}$,
where \siggeo\ is the geometrical cross section, $\bar{v}_{\rm e}$
is the electron thermal velocity, and \sigvin\ is the ion-neutral
momentum transfer rate coefficient (DRD83).}
The quantities $\nni-\nne$ and $\nni\vivec-\nne\vevec$
can be written in terms of the densities and velocities
of the other charged particles using the neutrality
condition, (\ref{neutrality_eq}), and the definition
of the total current,
\be
\jvec = \left(\nni\vivec-\nne\vevec+\ngp\vgpvec-\ngm\vgmvec
        +\nsgp\vsgpvec-\nsgm\vsgmvec\right)e.
\label{current_eq}
\ee
Making these substitutions in eq.~(\ref{sumdrft_eq}) we find,
after a little algebra, that
\begin{eqnarray}
\frac{\rhoi}{\tdragin}\left(\vnvec-\vivec\right) & = &
\frac{\left(\curl\Bvec\right)\cross\Bvec}{4\pi}
-\frac{e n_{{\rm g}^+}}{c}\left[\left(\vgpvec-\vivec\right)\cross\Bvec\right]
-\frac{e n_{{\rm sg}^+}}{c}\left[\left(\vsgpvec-\vivec\right)\cross\Bvec\right]  \nonumber \\
 & &
+\frac{e n_{{\rm g}^-}}{c}\left[\left(\vsgmvec-\vivec\right)\cross\Bvec\right] 
+\frac{e n_{{\rm sg}^-}}{c}\left[\left(\vsgmvec-\vivec\right)\cross\Bvec\right],
\label{plasma_eq}
\end{eqnarray}
where we used equation~(\ref{Efield_eq}) to express the
electric field in terms of \vivec\ and Amp\`{e}re's Law to
write the current in terms of $\curl\Bvec$.
Equation (\ref{plasma_eq}) says that the electromagnetic
and drag forces on the combined ion+electron fluid are in balance,
as they must be when ion and electron inertia are neglected.
Equation (\ref{plasma_eq}) is solved to find $\vivec(z,t)$ in terms of
the other fluid variables.

The ion energy equation is omitted in favor of the approximation
\be
\kB\Ti \approx \kB\Tn+\frac{1}{3}\mun\left|\vivec-\vnvec\right|^2
\label{Ti_eq}
\ee
(Chernoff 1987), which says that the ion fluid is in thermal balance
with heating and cooling due to ion-neutral scattering.

\subsubsection{Small Grains}

Our code solves the equations for mass conservation,
\be
\frac{\partial \rhosgm}{\partial t}
+
\frac{\partial}{\partial z}
\left(\rhosgm \vsgmz \right)
=
\Ssgm,
\label{sgm_mass_eq}
\ee
\be
\frac{\partial \rhosgz}{\partial t}
+
\frac{\partial}{\partial z}
\left(\rhosgz \vsgzz \right)
=
\Ssgz,
\label{sgz_mass_eq}
\ee
and
\be
\frac{\partial \rhosgp}{\partial t}
+
\frac{\partial}{\partial z}
\left(\rhosgp \vsgpz \right)
=
\Ssgp.
\label{sgp_mass_eq}
\ee
The source terms on the r.h.s. of equations
(\ref{sgm_mass_eq})-(\ref{sgp_mass_eq}) describe the
transfer of grains from one charge state to another
due to the capture of ions and electrons, and are
computed by integrating the appropriate cross sections
(Draine \& Sutin 1987) over the distribution of
relative velocities for charged particles and drifting
grains (see Appendix A).
We neglect the inertia of the small grains and
compute their velocities algebraically
by requiring the total force on each small-grain fluid to vanish.
Thus we require
\be
0 = \rhosgm\left(\vsgmvec-\vivec\right)\cross\Omegasgmvec
  + \frac{\rhosgm}{\tdragsgmn}\left(\vnvec-\vsgmvec\right)
  - \frac{\rhosgm}{\tcfsgmi}\vsgmvec
  + \frac{\rhosgz}{\tcfsgze}\vsgzvec ~~~,
\label{sgmdrift_eq}
\ee
\be
0 = \frac{\rhosgz}{\tdragsgzn}\left(\vnvec-\vsgzvec\right)
  - \frac{\rhosgz}{\tcfsgzi}\vsgzvec
  - \frac{\rhosgz}{\tcfsgze}\vsgzvec
  + \frac{\rhosgm}{\tcfsgmi}\vsgmvec
  + \frac{\rhosgp}{\tcfsgpe}\vsgpvec ~~~,
\label{sgzdrift_eq}
\ee
and
\be
0 = \rhosgp\left(\vsgpvec-\vivec\right)\cross\Omegasgpvec
  + \frac{\rhosgp}{\tdragsgpn}\left(\vnvec-\vsgpvec\right)
  + \frac{\rhosgz}{\tcfsgzi}\vsgzvec
  - \frac{\rhosgp}{\tcfsgpe}\vsgpvec,
\label{sgpdrift_eq}
\ee
where \Omegafvec\ ($=Q\Bvec/m_{\rm f} c$) is the cyclotron frequency for
particles of type f with charge $Q$ and \tcfij\ is the mean time
for a small grain to change its affiliation from fluid i to fluid j
by capturing an ion or electron.
As noted in previous studies (e.g., Pilipp et al.\ 1987;
Ciolek \& Mouschovias 1993), the capture of charged particles
by the grains transfers mass and momentum between the different
grain subfluids.
Our code includes these effects via the charge fluctuation
terms in the momentum equations and the source terms in the
mass equations.
The velocity dispersion of the small grains does not affect
other properties of the flow; consequently, we do not solve
the small-grain energy equation.

\subsubsection{Large Grains}

We neglect the velocity dispersion of the large grains
and omit the energy equation.
We include the large-grain inertia and solve the equations
for mass conservation,

%
%
\be
\frac{\partial \rhogm}{\partial t}
+
\frac{\partial}{\partial z}
\left(\rhogm \vgmz \right)
=
\Sgm,
\label{gm_mass_eq}
\ee
\be
\frac{\partial \rhogz}{\partial t}
+
\frac{\partial}{\partial z}
\left(\rhogz \vgzz \right)
=
\Sgz,
\label{gz_mass_eq}
\ee
\be
\frac{\partial \rhogp}{\partial t}
+
\frac{\partial}{\partial z}
\left(\rhogp \vgpz \right)
=
\Sgp,
\label{gp_mass_eq}
\ee
%
%
and momentum conservation,
\begin{eqnarray}
\frac{\partial}{\partial t} \left(\rhogm \vgmy \right)
+
\frac{\partial}{\partial z} \left(\rhogm \vgmy \vgmz \right)
 & = &
-\rhogm\left(\vgmz-\viz\right)\Omegag
+\frac{\rhogm}{\tdraggmn}\left(\vny-\vgmy\right) \nonumber \\
 &   &
-\frac{\rhogm}{\tcfgmi}\vgmy
+\frac{\rhogz}{\tcfgze}\vgzy, 
\label{gm_ymtm_eq}
\end{eqnarray}
\begin{eqnarray}
\frac{\partial}{\partial t} \left(\rhogz \vgzy \right)
+
\frac{\partial}{\partial z} \left(\rhogz \vgzy \vgzz \right)
 & = &
\frac{\rhogz}{\tdraggzn} \left(\vny - \vgzy \right)
+\frac{\rhogm}{\tcfgmi}\vgmy
- \frac{\rhogz}{\tcfgze}\vgzy 
\nonumber \\
& &
-\frac{\rhogz}{\tcfgzi}\vgzy 
+\frac{\rhogp}{\tcfgpe}\vgpy,
\label{gz_ymtm_eq}
\end{eqnarray}
\begin{eqnarray}
\frac{\partial}{\partial t} \left(\rhogp \vgpy \right)
+
\frac{\partial}{\partial z} \left(\rhogm \vgpy \vgpz \right)
 & = &
\rhogp\left(\vgpz-\viz\right)\Omegag
+\frac{\rhogp}{\tdraggpn}\left(\vny-\vgpy\right) \nonumber \\
 &   &
+\frac{\rhogz}{\tcfgzi}\vgzy
-\frac{\rhogp}{\tcfgpe}\vgpy,
\label{gp_ymtm_eq}
\end{eqnarray}
%
%
\begin{eqnarray}
\frac{\partial}{\partial t} \left(\rhogm \vgmz \right)
+
\frac{\partial}{\partial z} \left(\rhogm v_{{\rm g}^-z}^2\right)
 & = &
\rhogm\left(\vgmy-\viy\right)\Omegag
+\frac{\rhogm}{\tdraggmn}\left(\vnz-\vgmz\right) \nonumber \\
 &   &
-\frac{\rhogm}{\tcfgmi}\vgmz
+\frac{\rhogz}{\tcfgze}\vgzz,
\label{gm_zmtm_eq}
\end{eqnarray}
\begin{eqnarray}
\frac{\partial}{\partial t} \left(\rhogz \vgzz \right)
+
\frac{\partial}{\partial z} \left(\rhogz v_{{\rm g}^0z}^2\right)
 & = &
\frac{\rhogz}{\tdraggzn}\left(\vnz - \vgzz \right)
+\frac{\rhogm}{\tcfgmi}\vgmz
-\frac{\rhogz}{\tcfgze}\vgzz
 \nonumber \\
 &   &
-\frac{\rhogz}{\tcfgzi}\vgzz
+\frac{\rhogp}{\tcfgpe}\vgpz,
\label{gz_zmtm_eq}
\end{eqnarray}
and
\begin{eqnarray}
\frac{\partial}{\partial t} \left(\rhogp \vgpz \right)
+
\frac{\partial}{\partial z} \left(\rhogp v_{{\rm g}^+z}^2\right)
 & = &
-\rhogp\left(\vgpy-\viy\right)\Omegag
+\frac{\rhogp}{\tdraggpn}\left(\vnz-\vgpz\right) \nonumber \\
 &   &
+\frac{\rhogz}{\tcfgzi}\vgzz
-\frac{\rhogp}{\tcfgpe}\vgpz,
\label{gp_zmtm_eq}
\end{eqnarray}
where
$\Omegag \equiv
\left|\mbox{\boldmath$\Omega$}_{{\rm g}^-}\right|
=
\left|\mbox{\boldmath${\Omega}$}_{{\rm g}^+}\right|$.

\subsubsection{Chemistry}

The ion and neutral mass density depend on the chemistry,
which is calculated in parallel with the hydrodynamics.
Our code solves a set of chemical equations for the neutrals,
\be
\frac{\partial n_{{\rm n}\alpha}}{\partial t}
+
\frac{\partial}{\partial z}
\left(n_{{\rm n}\alpha}\vnz \right)
=
{\cal C}_{{\rm n}\alpha} - {\cal D}_{{\rm n}\alpha},    
\label{n_chem_eq}
\ee
and ions,
\be
\frac{\partial n_{{\rm i}\alpha}}{\partial t}
+
\frac{\partial}{\partial z}
\left(n_{{\rm i}\alpha}\viz \right)
=
{\cal C}_{{\rm i}\alpha} - {\cal D}_{{\rm i}\alpha},
\label{i_chem_eq}
\ee
where ${\cal C}_{{\rm f}\alpha}$ and ${\cal D}_{{\rm f}\alpha}$ are the
net rates per volume at which species $\alpha$ in fluid f is created and
destroyed, respectively.


\subsection{Validity and Relevance}

\subsubsection{Frozen Field Approximation}

The frozen field approximation is valid when $\Halli \gg 1$, where
\be
\Halli  = \frac{e B}{\mi c} \approx 36\,\frac{B_{-3}}{n_8}
\label{Hallival_eq}
\ee
is the ion Hall parameter,
$B=B_{-3}$\,mG is the magnetic field, and
$\nH \equiv 10^8n_8$\,\cmMMM.
The line-of-sight magnetic field components inferred from
Zeeman splitting in thermally-excited transitions
(summarized by Crutcher 1999) imply that $\Halli \gg 1$ for
$\nH \sim 10^7$\cmMMM\ but observations of denser regions are rare.
Extrapolating the thermal Zeeman data to higher densities
using the observed correlation $B \propto n^{\kappa}$ (with
$\kappa \approx 0.5$, see Crutcher 1999; Basu 2000) would imply that
the frozen-field approximation is valid for $\nH \simgt 10^9$\,\cmMMM.
The extrapolated field values are consistent with the
fields estimated in \water\ masers with $\nH \simgt 10^9$\,\cmMMM\ but
the field strengths inferred from maser observations are controversial
(Sarma, Troland \& Romney 2001 and references therein).

\subsubsection{Inertia of Ions and Small Grains}

Neglecting the inertia of ions and small grains is an
excellent approximation for {\it steady}\/ shocks.
Let \Ldragsg\ be the length scale for gas drag to decelerate
the small grains, \Lneut\ the scale for compression of the
neutrals, and \Lfield\ the scale for compression of the magnetic field.
Figure~2 compares these and other length scales for a shock
with speed $\vs=10~\kms$ and various preshock densities.
The dynamical scales were estimated using dimensional analysis
of the equations of motion for a steady shock
(DRD83; see eq.~[3]--[4]):
\be
\Lneut = {\rho_{\rm n0} \, v_{\rm s}^2 \over F_{\rm n} },
\label{Lneut-eq}
\ee
and
\be
\Lfield = {B_0^2 \over 8\pi F_{\rm i}},
\label{Lfield-eq}
\ee
where the subscript ``0'' denotes preshock quantities.
We took the preshock magnetic field to be
\be
B_0 = b\,\left(n_0/\cc\right)^{1/2}\,\mu{\rm G}
\label{Bzero_eq}
\ee
with $b=0.3$.
We included the drag due to electron-neutral and ion-neutral
scattering (DRD83, eq.~[11]-[14]) as well as the gas drag on
small and large grains (D80, eq.~[40]).
We computed \Lfield\ at a typical point in the magnetic precursor
where the neutral particles are unperturbed and the magnetic
field has been compressed by a factor of 2 (see D80, eq.~[19]).
For \Lneut\ we took a representative downstream point, with
unperturbed neutrals and the magnetic field compressed by a
factor of $\Mna$, the neutral Alfv\'{e}n Mach number
(see Kaufman \& Neufeld 1996a and references therein).
Preshock chemical and grain abundances were taken from Figure~1
and the drift velocities of large and small grains were computed
by neglecting their inertia for expediency.
We set
\be
\Ldragsg = v_{{\rm sg}^{-}z} \, \tdragsgmn,
\label{Ldragsg_eq}
\ee
with the grain velocity evaluated in a frame where
the shock is steady and the drag time
computed from the velocity-dependent drag
force on a drifting sphere (Draine \& Salpeter 1979).
>From Figure~2 and analogous calculations for other shock speeds
we conclude that the inertia of small grains (and hence
the ions) can be neglected in steady, nondissociative shocks
with $\vs<30~\kms$.

In {\it time-dependent}\/ shocks, the inertia of light particles
is a subtler issue.
The formation of a magnetic precursor depends on the speed
at which compressive disturbances are communicated to the charged fluid,
and this ``signal speed'' is modified when the inertia of
light particles is neglected.
As a simple illustration, consider the propagation of small-amplitude
waves in a plasma of ions, electrons, and neutrals.
In Figure 3 we plot the phase velocity of waves
propagating perpendicular to \Bvec, as predicted by the exact
dispersion relation for a cold plasma (solid curves) and the
dispersion relation with ion inertia neglected (open circles).
The exact dispersion relation predicts the well-known
result (Kulsrud \& Pearce 1969) that compressive waves propagate
in the ion-electron fluid for wavelengths
\be
\lambda < \lambda_1 = 4\pi\vims\tdragin
\label{lambda1_eq}
\ee
and in the bulk plasma for
\be
\lambda > \lambda_2 = \lambda_1\,\sqrt{\rho_{\rm n}/4\rho_{\rm i}},
\label{lambda2_eq}
\ee
where $\vims=B/\sqrt{4\pi\rhoi}$ is the ion magnetosound speed
for a cold plasma.
When the ion inertia is neglected, the short-wavelength mode
disappears and the fastest ``signal speed'' for propagating waves drops
from \vims\ to the bulk magnetosound speed,
$\vms \approx \vims\,\sqrt{\rhoi/\rhon}$.
This seems to suggest that the ``inertialess equations'' would fail to
admit C type solutions for $\vs > \vms$.
However Smith and Mac Low (1997) have shown,
by direct numerical integration of the inertialess equations,
that a J shock with $\vs>\vms$ can evolve into a C shock.
We describe a similar result in \S3.
The development of a magnetic precursor in both examples
is initiated by rapid ion motions with $\vi\sim\vims$.
Evidently the formation of a C shock does not require {\it propagating}\/
waves with phase velocities larger than the shock speed.
We speculate that the ions are accelerated
by evanescent waves in the ``forbidden zone''
$\lambda_1 < \lambda < \lambda_2$, which have long
damping times (Fig.~4).
If correct, our conjecture may have interesting implications for
maximum speed of a C shock.
This and related issues will be discussed elsewhere
(Roberge \& Ciolek, in preparation).

\subsubsection{Relevance of Large-Grain Inertia and Charge Fluctuations}

The drag length for large grains [$L_{\rm drag}$(g)] is shown in
Figure~2.
For $\nH \simlt 10^6$\,\cmMMM, the large grains are partially
coupled to \Bvec\ and may affect the formation of C shocks
by ``loading'' the field lines (Draine \& McKee 1993; see \S1).
Over this density range the drag length is comparable
to \Lfield\ (Fig.~2b), so including the large-grain inertia is
necessary.
Length scales for the capture of charged particles by small
(sg$^0$+e, sg$^-$+i) and large (g$^0$+e, g$^-$+i) grains are
also plotted in Figure~2.
The charge fluctuation times were computed using
abundances from Figure~1 and rate coefficients from Appendix~A.
Because the shock structure is sensitive to changes in
the ionization (Flower, Pineau des Forets \& Hartquist 1985;
Pilipp et al.\ 1990), it is necessary to include the
charge fluctuations of small grains for
$\nH \simlt 10^9$\,\cmMMM\ (Fig.~2a).
Figure~2b shows that large grains undergo many charge fluctuations
as the magnetic field is compressed, partially coupling the
neutral grains to the field lines. 
Consequently the charge fluctuations of large grains are
also important.


\section{Solutions}

\subsection{Initial and Boundary Conditions}

We model two types of problem with our code.
In the first type (``free propagation problems''),
the initial conditions are some disturbance, linear or nonlinear,
specified on a finite computational domain, $0<z<Z$.
For free propagation problems we impose ``non-reflecting
boundary conditions'' (discussed below) at $z=0$ and $z=Z$.
These boundary conditions allow the disturbance to propagate
off the domain as if the boundaries were at infinity.
In the other type of problem (``inflow problems''),
the initial conditions are a uniform, static ``undisturbed state'' for
$0<z<Z$ and a steady flow is incident at the boundary $z=0$.
For inflow problems we specify the flow variables at $z=0$
and impose nonreflecting boundary conditions at $z=Z$.

For inflow problems, the undisturbed state is specified by
choosing the density $n_0$ and magnetic field strength, \Bo.
In the exemplary calculations presented in this paper,
the temperature of the undisturbed gas is not of interest.
We assign the initial neutral, ion, and electron
temperatures to reasonable but somewhat arbitrary values
instead of, say, requiring the fluids to be in thermal balance.
Once the gas density and temperature are specified,
the initial densities $n_{{\rm f}\alpha 0}$ of the various
chemical species are determined by solving the chemical rate equations
for a static system (see \S2.2.7).
To determine the abundances uniquely we must
specify the fractional abundance of heavy nuclei
(\xmetal), and the initial abundances of large (\xgo), and
small (\xsgo) grains.
We typically set $\xmetal \sim 10^{-6}$.
The initial grain abundances are specified in practice
by choosing the grain radii, \ag\ and \asg,
the densities \rhog\ and \rhosg\ of the grain materials, 
and the initial dust-to-gas mass ratios, \chigo\ and \chisgo.
For diffuse regions of molecular clouds or protostellar cores,
we typically set $\ag=0.1$\,\micron\ $\rhog=3.3$\,\gcmMMM,
and $\chigo = 10^{-2}$, yielding $\xg = 2 \times 10^{-12}$.
For the small grains we usually take
$\asg=4$\,\AA, $\rhosg=2.2$\,\gcmMMM, and $\chisgo=5 \times
10^{-5}$, yielding an abundance, $\xsgo=2\times 10^{-7}$,
consistent with the constraints on small grains in diffuse gas
(Puget \& Leger 1989; Weingartner \& Draine 2001; Li \& Draine 2001).
For models of shocks in dense gas, we adjust the grain radii
and dust-to-gas ratios to account approximately for the
possible depletion of small grains, coagulation of large grains,
(e.g., Ossenkopf 1993), and depletion of large grains due to
ambipolar diffusion during core formation
(Ciolek \& Mouschovias 1994, 1996).

The actual problem we solve is a dimensionless one. The
governing equations (\S~2) are put into nondimensional
form by using as a unit of distance some specified length $L$,
with $Z = {\cal R} L$, ${\cal R}$ a constant. Velocities
are normalized to $\Czero$, the isothermal sound speed in
the undisturbed neutral gas. (For free propagation problems,
``undisturbed gas'' means the gas outside the initial disturbance.)
The unit of time is therefore $L/\Czero$.
Temperatures and densities are normalized to the temperature,
\Tno, and mass density, \rhono, of the undisturbed neutral gas, and
the unit of magnetic field strength is taken to be $\Bo$.
A consequence of our selection of units is that the
dimensionless Alfv\'{e}n speed of the undisturbed
gas, $\vano/\Czero$ [$=\Bo/(4 \pi \rhono)^{1/2} \Czero$],
appears as a characteristic parameter in the problem. 

The use of non-reflecting boundary conditions
(also referred to as ``transmissive'' or ``radiation'' boundary 
conditions) allows the flow to pass unimpeded through the
computational boundaries, and prevents the reflection of
spurious waves from the boundaries back into the computational domain.
A comprehensive review of non-reflecting boundary conditions in
numerical hydrodynamics is presented by Givoli (1991). 
We use the following formulation, due to Thompson (1987)
and Vanajakshi, Thompson, \& Black (1989). 
Examination of the governing equations (\S2)
reveals that they are all conservative transport equations of the form
\begin{equation}
\label{nonrefleqa}
\frac{\partial \Uvec}{\partial t} + \frac{\partial \Fvec}{\partial z}
+ \Cvec = 0,
\end{equation}
where
\begin{equation}
\Uvec= \left[
\begin{array}{cl}
\rhon \\
\rhon \vny \\
\rhon \vnz \\
\frac{1}{2}\rhon v_{\rm n}^2 + \En \\
\vdots \\
\end{array}
\right]
\end{equation}
is the (column) vector of conserved variables,
$\Fvec$ is the corresponding vector flux of conserved quantities, and
$-\Cvec$ is the vector of source terms.
The system (\ref{nonrefleqa}) is transformed into the quasilinear form
\begin{equation}
\label{nonrefleqb}
\frac{\partial \Uvecb}{\partial t} + \Amatrix \cdotm \frac{\partial \Uvecb}{\partial z}
+ \Cvecb = 0,
\end{equation}
where 
\begin{equation}
\Uvecb = \left[
\begin{array}{c}
\rhon \\
\vny \\
\vnz \\
\En/\rhon \\
\vdots
\end{array}
\right] ~~,
\end{equation}
is the vector of primitive variables and the Jacobian matrix,
$\Amatrix(\Uvecb)$, is readily determined. The eigenvalues
$\left\{\xi_k\right\}$ of $\Amatrix$ are the characteristic speeds of
the flow. Non-reflecting boundary conditions are implemented by imposing
the set of equations 
\begin{equation}
\label{nonrefleceqc}
\lvec_{k} \cdotm \frac{\partial \Uvecb}{\partial t} + {\cal L}_{k} +
\lvec_{k} \cdotm \Cvecb = 0,
\end{equation}
at the appropriate boundary,
where $\lvec_{k}$ is the left (row) eigenvector associated with the 
eigenvalue $\xi_{k}$ and
\begin{equation}
\label{Lieq}
{\cal L}_{k} =
\left\{
\begin{array}{c}
\xi_{k} \lvec_{k} \cdotm \partial \Uvecb/\partial z  \hspace{3em}
\mbox{for outgoing waves,}\\ \\
\hspace{2em} 0 \hspace{6em} \mbox{for incoming waves.}
\end{array}
\right.
\end{equation}
As demonstrated by Thompson (1987) and Vanajakshi et al. (1989), the
conditions (\ref{Lieq}) for outgoing waves ($\xi_{k} < 0$ at $z=0$,
$\xi_{k} > 0$ at $z=Z$) and incoming waves ($\xi_{k} > 0$ at
$z=0$, $\xi_{k} < 0$ at $z=Z$)
eliminate spurious reflections and produce accurate numerical
solutions at the artificial boundaries.

\subsection{Numerical Method}
The dimensionless governing equations with nonreflective and/or
inflow boundary conditions comprise a set of coupled, nonlinear,
partial differential equations (PDEs).
We solve the PDEs numerically
using a technique known as the ``method of lines" (e.g., Liskovets 1965;
Schiesser 1991).
The computational domain is divided into $N$ cells of
uniform width $\Delta z = Z/N$ with cell boundaries at
$\{z_j=j\Delta z,~j=0,1,...,N \}$.
Mean values of the variables are assigned to the
center of each cell, i.e., to the points
\begin{equation}
\label{zbareq}
\zbar_{j} \equiv \frac{(z_{j}+z_{j-1})}{2}=z_{j-1}+\frac{1}{2}\dz
=(j-\frac{1}{2})\dz,
\end{equation}
for $j\geq 1$.
The mean value of variable $f$ in cell $j$ will be denoted by
$\fbar_{j} \equiv f(\zbar_{j})$.
In the method of lines,
spatial derivatives
are approximated using finite differences,
thereby converting the PDEs into a system of ordinary differential
equations (ODEs) of the form
\begin{equation}
\frac{d \yvec}{d t} = \fvec[\yvec(t),t],
\end{equation} 
where $d\yvec/dt$, $\yvec$, and $\fvec$ are vectors of length
$WN$, and $W$ is the number of variables defined at each cell.
The ODEs are solved using standard stiff ODE solving subroutines readily
available in the public domain.

For cells inside the computational domain, spatial derivatives are
approximated with the fourth-order differencing scheme 
\begin{equation}
\label{gradapproxeq}
\left[\frac{\partial f}{\partial z}\right]_{j} \doteq 
\frac{-(1/12)(\fbar_{j+2} - \fbar_{j-2}) 
+ (2/3)(\fbar_{j+1}-\fbar_{j-1})}{\dz}~~~.
\end{equation}
For cells adjacent to the boundaries, we use the three-point scheme
\begin{eqnarray}
\label{gradapproxeqja}
\left[\frac{\partial f}{\partial z}\right]_{1} &\doteq& 
\frac{(1/3)f_{2} + f_{1} - (4/3)f_{b0}}{\dz}~~~, \\
\label{gradapproxeqjb}
\left[\frac{\partial f}{\partial z}\right]_{N} &\doteq& 
\frac{(4/3)f_{bN} - f_{N} - (1/3)f_{N-1}}{\dz}~~~,
\end{eqnarray}
where $f_{b0}$ and $f_{bN}$ refer to the values of variables
at the boundaries $z=0$ and $z_{N}=Z$, respectively.
Thus, derivatives at the boundaries depend only on variables inside the
computational domain. They are accurate to $\Order(\dz)^2$, while the
interior derivatives (eq. [\ref{gradapproxeq}]) are accurate to
$\Order(\dz^4)$.

Advection equations, of the form
$\partial f/\partial t = -\partial(f v)/\partial z$,
are discretized using the monotonic, upwind algorithm of van Leer
(1979). Following this algorithm, advection terms are
approximated as
\begin{equation}
\label{advectapproxeq}
\left[\frac{\partial(fv)}{\partial z}\right]_{j} \doteq - \frac{(\fR\vR - \fL\vL)}{\dz}~~~,
\end{equation}
where the flux $\fR\vR$ at the right cell face is approximated by
setting $\vR=(\vbar_{j}+\vbar_{j+1})/2$, and 
\begin{equation}
\label{fReq}
\fR=
\left\{
\begin{array}{c}
\fbar_{j} + \frac{1}{2}\delfbar_{j} \hspace{4em}{\rm{if}}~ \vR \geq 0, \\
\\
\fbar_{j+1} - \frac{1}{2}\delfbar_{j+1} \hspace{2em}{\rm{if}}~ \vR < 0,
\end{array}
\right.
\end{equation}
and the flux $\fL\vL$ at the left cell face is approximated
by setting
$\vL=(\vbar_{j-1}+\vbar_{j})/2$ and 
\begin{equation}
\label{fLeq}
\fL =
\left\{
\begin{array}{c}
\fbar_{j-1} + \frac{1}{2}\delfbar_{j-1} \hspace{2.5em}{\rm{if}}~ \vL \geq 0, \\ \\
\fbar_{j} - \frac{1}{2}\delfbar_{j} \hspace{4.5em}{\rm{if}}~ \vL < 0.
\end{array}
\right.
\end{equation}
Here we have introduced the ``monotonic derivative,''
\begin{equation}
\label{delfbareq}
\delfbar_{j}=
\left\{
\begin{array}{cl} 
 \frac{(\fbar_{j+1} - \fbar_{j})(\fbar_{j}-\fbar_{j-1})}{\fbar_{j+1}-\fbar_{j-1}}
\hspace{2em} {\rm{if}}~(\fbar_{j+1}-\fbar_{j})(\fbar_{j}-\fbar_{j-1}) > 0, \\
\\
\hspace{-5em} 0 \hspace{5.5em} \rm{otherwise}.
\end{array}
\right.
\end{equation}
For the boundary cell adjacent to $z=0$ we set $\vL = v_{b0}$.
If $\vL \geq 0$, $f_{L} = f_{b0}$, otherwise $f_{L}$ is given by
equation (\ref{fLeq}). Similarly, for the cell adjoining $z=Z$,
$v_{R} = v_{bN}$. If $v_{R} \geq 0$, equation (\ref{fReq}) is used for
$f_{L}$; if $v_{R} < 0$, $f_{R} = f_{bN}$ instead.

\subsection{Benchmark Tests}

To establish the accuracy of our numerical code, in this section we
present the results of various test runs and compare them with
known solutions.
In all of the tests presented here, we set
$\En = \Pn/(\gamma - 1)$, where $\gamma$ is the adiabatic index
of the neutral gas, assumed constant.
In models that include cooling, we use the
analytical cooling function $\Ln$ defined by Roberge \& Draine (1990,
eq. [3.18]). 

\subsubsection{Shock Tube}

A standard test for numerical codes is the well-known shock tube
problem (e.g., Sod 1978). In the absence of magnetic fields, the
solution for this type of Riemann problem can be found exactly
using analytic methods (Toro 1997 and references therein). 
The gas is adiabatic in the shock tube test, with
$\gamma = 5/3$. The magnetic field strength is made vanishingly
small; in this limit, the ions, electrons, and grains move with
the neutrals (due to collisional forces) and do not appreciably affect
the flow, because the density of charged particles
is much smaller than the neutral gas density.

The shock tube test falls into the category of ``free propagation''
problems with nonreflecting boundary conditions at both edges
of the spatial domain (see \S3.1).
The neutral gas is initially at rest.
The initial density and temperature profiles are constant on
both halves of the domain with a discontinuity at $z=Z/2$.
In our test we chose
$\rhon(z) = 12 \rhono$ and $\Tn(z) = 1.875 \Tno$ for $z < Z/2$
and $\rhon(z) = \rhono$ and $\Tn(z) = \Tno$ for $z > Z/2$,
where \rhono\ and \Tno\ are scaling values that merely define
the dimensionless units (see \S3.1).

For this test we turned off the source terms in the continuity
equations for all fluid species. Because the gas flow is taken
to be adiabatic, the cooling function for the
neutral gas is set to zero.

The results of the shock tube test for a dimensionless
time $t_{1} > 0$ are
displayed in Figures $4a$-$4d$ (circles).
This particular model used 400 mesh
points; for visual clarity in each plot we display only every eighth
data point. Figure $4a$ shows $\rhon(z,t_{1})/\rhono$ as a function of
$z/Z$. The ratio $\vnz(z,t_{1})/\Czero$ is exhibited in Figure $4b$.
The pressure $P_{\rm n} (z,t_{1})$ relative to $P_{\rm{n,0}}$, and
$T_{\rm n}(z,t_{1})/T_{\rm{n,0}}$ are shown in Figures $4c$ and $4d$,
respectively. For comparison, also displayed in these figures are the
exact Riemann solutions (solid curves).
Examination reveals that our multifluid code effectively reproduces the
full structure of the shock tube problem, including the shock wave,
rarefaction wave, and contact discontinuity which are present at this
time. The r.m.s. relative errors of the numerically-computed variables
are typically $ \simlt 10^{-2}$
--- except at the oscillations near the base of the rarefaction
wave and at the contact discontinuity.

Figures $5a$-$d$ present numerical results for the same model run 
(circles) at a later time $t_{2}$, which are again compared to the
exact solution (solid curves).
By this time, both the forward-moving shock wave and contact 
discontinuity, and also the backward propagating rarefaction wave, all
present in Figures $4a$-$d$, have moved ``off stage", that is, past
the boundaries at $z = Z$ and $0$. We note that passage of the forward
waves through $z=Z$ does not produce any
substantial numerical reflections at the boundary located there. Nor are
there any spurious reflections generated by the backward
rarefaction wave incident upon the opposite boundary at $z=0$. This
illustrates the efficacy of the non-reflecting boundary conditions
(\S~3.1).

\subsubsection{Nonmagnetic Shock with Inflow Conditions}

The next test we present is an inflow problem.
We again consider nonmagnetic, adiabatic ($\gamma = 5/3$) flow.
Initially the neutral gas within the entire region
[0, $Z$] is at rest with uniform density \rhono\ and
temperature \Tno.
At $t=0$ a high-velocity shock with $\vnz = 50 \Czero$ is incident on the
boundary at $z=0$; this corresponds to a Mach number
$\machno = \vshk/\gamma^{1/2}\Czero = 51.7$, where $\vshk$ is the
shock speed.
The inflowing shocked material has density $\rhon = 3.99 \rhono$,
and temperature $\Tn = 835 \Tno$. 

Figures $6a$-$d$ depict the profiles of the density, velocity, pressure,
and temperature of the neutrals, normalized to the uniform initial
state, at a time $t > 0$. As before, the circles represent the solution
from our multifluid numerical code. The solid line is the exact
Riemann solution. Comparison indicates that there is good agreement
between the numerical and exact solutions. 

\subsubsection{Counterpropagating Rarefaction Waves in a Magnetic,
Flux-Frozen Fluid}
This test is used to verify that our numerical code has the correct
MHD behavior in the limit of flux-freezing, that is, when the
magnetic field is frozen into the bulk fluid. Effective freezing of the
magnetic flux into the neutrals can occur when the characteristic 
minimum flow time $\tau_{\rm{flow,min}}$ ($= \dz/\vnz$) is
$\gg \tau^{\rm d}_{\rm{n,tot}}$,
the drag time for neutrals colliding with all charged particles
attached to the magnetic field.
In this situation, information about the magnetic field is quickly
communicated to the neutrals, through collisions with the plasma, on a
time scale much smaller than the natural time scale associated with the
flow. When this happens, the magnetic field, plasma, and the
neutrals can rapidly respond to one another, and thus move together as a
single combined fluid. In our code, the condition
$\tau_{\rm{flow,min}}/\tau^{\rm d}_{\rm{n,tot}} \gg 1$ is readily
satisfied by using a state with a large enough abundance of charged
particles (since $\tau_{\rm{n,f}} \propto x_{\rm f}^{-1}$; e.g., see
eqs. [2c] and [10a] of Ciolek \& Mouschovias 1993), and/or
choosing a sufficiently large value $Z$ for the size of the physical
region being modeled (recall that $\dz = Z/N$). 

The model we choose for this test is the counterpropagating MHD
rarefaction flow described in \S~4 of Ryu \& Jones (1995).
This test consists of two oppositely-directed, adiabatic
rarefaction waves in a magnetized fluid. The initial state has a
nonzero, discontinuous velocity profile with $\vnz(z < Z/2) = -\Czero$,
and $\vnz(z > Z/2) = \Czero$. However, all of the other physical
quantities are uniform in the domain [$0,Z$]:
$\rhon(z) = \rhono$, $\Tn(z)=\Tno$, and $\Bx(z)=\Bo$.
The Alfv\'{e}n speed in the undisturbed neutrals, $\vano$,
is equal to $\Czero$ for this model.
Creation/destruction of charged species was not considered in the
analysis of Ryu \& Jones. Hence, to be able to make an accurate
comparison to with their published model results, in this test we
switched off the source and sink terms that appear in the continuity
equation for each charged species.

Figures $7a$-$d$ exhibit the state of this test model at a time $t > 0$,
when the forward- and backward-traveling rarefaction waves have
opened a cavity in the density and magnetic field where mass and magnetic
flux have flowed away from the center.
The curves with circles show ({\it a}) the density, ({\it b}) velocity,
({\it c}) pressure, and ({\it d}) the magnetic field profiles generated
by our numerical code at this particular time. For comparison, also
shown in each plot is the solution (solid lines) of Ryu \& Jones (1995,
their Fig.  3b), which they obtained both numerically and by using a
specially-developed MHD Riemann solver. These figures show that our
multifluid code reproduces the limiting ideal MHD flow. That the flow
in this model is indeed
effectively flux-frozen can be seen by comparing Figures $7a$ and $7d$,
which show that the mass and magnetic flux have behaved in the same
way, preserving the local mass-to-flux ratio
$(\rhon/\Bx)/(\rhono/\Bo) = 1$ throughout the evolution. There is, however, a numerical overshoot at the half-way
point, $z = Z/2$, which is the location of the contact discontinuity within
the initial state. A similar overshoot was also found to occur in the
numerical simulation of Ryu \& Jones.  

Finally, we note that there are no spurious oscillations at
either boundary in Figures $7a$-$d$. This confirms that the
non-reflecting condition imposed at the artificial boundaries in our
code is also properly dealing with the passage of MHD flows.

\subsubsection{Chernoff's Solution for Adiabatic C Shocks}

We also test the effectiveness of our code in dealing with
realistic, non-ideal MHD situations, such as when the magnetic field is
not necessarily frozen into the neutrals, due to relatively small
abundances of charged particles. To examine this regime, we modeled
the formation of a C shock in a weakly-ionized plasma. As noted
earlier, this problem for three-component plasmas (neutrals, ions, and
electrons) has previously been considered in great detail for both
steady (e.g., D80; DRD83)
and also nonsteady (e.g., T\'{o}th 1994; Smith \& Mac Low 1997)
flows.

For the case of time-independent flow, a detailed phase plane analysis
of `energy-conserving' (i.e., adiabatic) MHD shocks was presented by
Chernoff (1987) for three-component plasmas. We use his results as the
benchmark for this particular test.

To compare our results to those of Chernoff, we used a model
containing relative abundances of dust grains that are orders of
magnitude smaller than the ``typical values'' normally adopted
(\S~3.1). This reduced the effects of the dust grains,
making our test model effectively a three-component plasma. We also
neglected the source terms in the 
continuity equations, since chemistry was also
neglected by Chernoff. The source terms for momentum transfer due to
collisions between ions and neutrals were taken from DRD83 and
Ciolek \& Mouschovias (1993).

For initial conditions, we used an adiabatic inflow model
with a Mach number $\machno=9.75$ incident on a uniform
initial state. The values for the density, temperature, and magnetic
field strength of the inflow at $z=0$ were determined by the jump
conditions for a perpendicular, ideal (i.e., frozen-in magnetic flux)
MHD  J shock (e.g., Priest 1984). The Alfv\'{e}n speed in the neutrals
for the pre-shocked initial state is $6.17 \Czero$.

As the flow expands into the computational region for $t>0$, the initial
J shock rapidly evolves into a C shock. The transient, short-term
evolution during the transition from a J-type to C-type shock is
similar to that which was reported by T\'{o}th (1994), and Smith
\& Mac Low (1997), including an early, rapid diffusion of the ions and
magnetic field outward from the initially constricting J shock (with
$\viz \gg \vnz$ in this phase of the evolution --- for more discussion
of transient phenomena in a representative model, see \S~3.3.5
below).

A steady C shock is soon established, with a velocity
$\vshk = 12.6 \Czero$. The Alfv\'{e}n Mach number
$\Mna$ ($= \vshk/\vano$) for this shock is 2.04; this
result is consistent with the analysis of Chernoff (1987), who found
that under these conditions (`energy conserving' flow), C shocks should
exist for $\Mna < 2.8$.
Figure $8a$ shows the profile of the neutral velocity (circles) and the
ions (filled diamonds) at a time after the steady state is attained.
The smooth, supersonic-to-supersonic transition in the
flow, characteristic of a C shock, is apparent in this figure, and so is
the magnetic precursor in the ion fluid that precedes the
neutral flow at the head of the shock.

The earlier studies of MHD shocks in partially ionized fluids
were usually carried out in the frame of reference moving with the
shock. Figure $8b$ displays the shock-frame velocities
of the neutrals ($\vnzsf=\vnz - \vshk$; circles), and 
ions ($\vizsf=\viz - \vshk$; filled diamonds), for
the same time as in Figure $8a$. As in
previous examples, the velocities are normalized to $\Czero$.

We may also quantitatively compare our numerical results to the
detailed predictions of the phase-plane analysis of shocks in
three-component plasmas by Chernoff (1987). Figure
$8c$ displays the values (circles) of the phase variable
$r = |\vnzsf|/\vshk$ as a function of $q = |\vizsf|/\vshk$, for
the numerical results presented in Figure $8b$. The dashed lines in Figure
$8c$ are the analytical values for $r(q)$ calculated by using
equation (35) of Chernoff (1987), for a shock with an Alfv\'{e}n
Mach number $\Mna = 2.04$. The essential equality of these two
curves in this figure indicates that the results of our multifluid
numerical code are in accord with the analytical solution. 

\subsubsection{Illustrative Solution with Grain Dynamics}

We now present an exemplary model that demonstrates some of the
unique features of our time-dependent, multifluid numerical 
MHD code. As described above in \S~2, this includes the ability to
follow non-steady, transient phenomena, as well being able to follow
simultaneously the dynamics of several different fluid species.

The illustrative model we present is an inflow J shock. The 
flow for $z < 0$, $t< 0$, is assumed to have a frozen-in magnetic
field (so that the plasma, magnetic field, and neutrals move together),
with the velocity and other quantities at $z=0$ 
determined by the jump conditions for a perpendicular shock.
The resulting flow inside the computational domain $[0,Z]$ is
then determined by solving the full set of governing equations,
including the continuity equations for all 9 fluids (with source
terms), and also the analytical cooling function $\Ln$ (\S~3.3).

For specificity, the pre-shock neutral gas has a uniform number density
$\nno = \rhono/\mun = 1.03 \times 10^8~\cc$, and temperature
$\Tno = 10$ K. The magnetic field of the uniform initial state
in this model is $\Bo = 600~\mu{\rm G}$. The initial abundances of
the various particle species are $\xxio = 2.04 \times 10^{-10}$,
$\xxeo = 3.26 \times 10^{-12}$,
$\xgmo= 4.13 \times 10^{-13}$, $\xgzo=2.34 \times 10^{-12}$,
$\xgpo=3.42 \times 10^{-13}$, $\xsgmo=9.94 \times 10^{-10}$,
$\xsgzo=9.82 \times 10^{-8}$, and $\xsgpo=7.93 \times 10^{-10}$.
The density and magnetic field strength of this state are such that
the Hall parameter for the small charged grains is
$\Gamma_{\rm{sg}} = \Omegasg \tau^{\rm d}_{\rm{sg,n}} = 25.4 \gg 1$.
This means that the magnetic field in this model is effectively frozen
into the plasma of ions and also the small charged grains.
The Hall parameter of the large grains is
$\Gamma_{\rm g} = \Omegag \tau^{\rm d}_{\rm{g,n}} = 2.54 \times 10^{-3} \ll 1$,
so they are not attached to the magnetic field; this means
that their dynamics are primarily governed by collisions with the
neutrals. 
Finally, the inflowing J shock has $\machno = 40$, and $\Mna=115$,
corresponding to a speed $\vshk=9.8~\kms$.

We first examine the model flow on a length scale $\sim 10^{14}$ cm.
On this length scale, and for these parameters, the shock travels as a
typical J shock. Figures $9a$-$9c$ shows the profile of various physical
quantities at $t=5.5 \times 10^8$s. Over these length and time scales,
the coupling between the neutrals and the other species is rapid enough,
compared to the characteristic flow time of the shock, that the
overall motion of the fluids is essentially as a single, combined
fluid.

In contrast, we now consider the same model on the much smaller scale
$\sim 10^{11}$ cm. This scale is now small enough that the effect
of drift of the magnetic field and the plasma with respect to the
neutrals is discernible. So too, is the drift or lag of the large
dust grains with respect to the neutrals because of inertial
effects. This is demonstrated in Figures $10a$ and $10c$, which
show the velocities of the neutrals (circles), ions and small charged
grains (filled diamonds), and the large dust grains (crosses) at times
$t = 1.32 \times 10^3$s and $t=3.68 \times 10^3$s, respectively.
As can be seen from these figures, these times are so early on
that very little of the inflowing neutral gas has penetrated into
the physical/computational region. 
However, there has been significant
readjustment of the ions and the small charged grains, due to plasma
drift. The diffusion of the charged particles is now resolved; 
the plasma and magnetic field have become `unlatched' from the inflowing
neutrals, as evident by the higher-velocity precursor in these figures.

Figure $10b$ shows the profiles of the neutral density (circles)
and the magnetic field (filled diamonds), normalized to the uniform
reference values, at $t=1.32 \times 10^3$s; Figure $10d$ shows the
same thing at $t=3.69\times 10^3$s. Again, there is little penetration
of the neutrals into the computational region during this early phase
of the evolution.
Because the magnetic gradient is much steeper at earlier
times (being greatest at time $t=0$, when the gradient is initially the
discontinuity of the J shock), the velocity of the ions and the small
charged grains is much larger at earlier times. This accelerates the
plasma ahead of the shock, and gives rise to the precursor.
The driving gradient decreases with time as the
magnetic field spreads out into the lower-density, lower-magnetic field
strength undisturbed region ahead of the shock, and the velocity of the
plasma correspondingly decreases from its maximum. From Figures $10a-d$
the effective signal or disturbance speed of the precursor at these
early times is found to be $\sim 10^2~\kms$. This value is intermediate
between the ion magnetosound speed and the magnetosound speed for the small
charged grains, which are $1800~\kms$ and
$36~\kms$, respectively.  

Notably lagging behind the neutrals in Figures $10a$ and $10c$ are the
large dust grains; only a few non-zero velocity data points for the
large grains (crosses) appear in the lower left-hand corners of these
figures. This is because of the inertia of the large grains,
which is noticable on these scales. The distance over which the large
grains are accelerated due to collisions with the neutrals is
$\Ldragg =  (\vnz - v_{{\rm g},z}) \tau^{\rm d}_{\rm{g,n}} = \rho_{\rm s}\ag/\mun\nn$,
where we have used the previously defined collision time scale, the
relation $\mg = (4 \pi/3)\rho_{\rm s} \ag^{3}$, and the fact that the
relative drift speed in the shock is much greater than the sound speed.
The neutral density in the shock front is $ \simeq 4 \nno$,
and using $\mun = 2.33$ amu for the assumed $\Htwo$ gas with a
10\% abundance of He by number, we therefore have
$\Ldragg \approx 0.3 \times 10^{11}$cm. Hence, the lag of the
large grains behind the neutrals on these length scales is understood
for this model.

The state of this model at $t= 2.63\times 10^5$s is shown in
Figures $11a$ and $11b$, and also at $t= 5.5 \times 10^5$s, in
Figures $11c$ and $11d$. By these times the adjustment by plasma drift is
approaching an asymptotic state, with an ion and small-charged grain
mediated magnetic precursor still ahead of the shock. A particularly
interesting feature in the magnetic field profile can be seen in
Figures $11b$ and $11d$, which show that, because of the diffusion
of the plasma with respect to the neutrals, magnetic field has
``leaked forward" ahead of the shock. This decreases the magnetic field
strength $\Bx$ immediately behind the shock front relative to the far
downstream (frozen-in) value $ \simeq 4 \Bo$.

The lag of the large dust grains (crosses) behind the neutrals (circles)
is again apparent in Figures $11b$ and $11d$. The distance over which
the motion of the large grains is delayed with respect to the neutrals,
because of large grain inertia, is consistent with, and explained by,
the value for $\Ldragg$.

Finally, we point out that the results presented in Figures $9a$-$9c$
are entirely consistent with the model results shown in Figures
$10a$-$10d$ and $11a$-$11d$. This is so because the small-scale structure
has a maximum extent that is basically the size of the magnetic
precursor, which has a width $\simlt 10^{12}$ cm. However,
examination of Figures 10 and 11 reveal that on scales greater than this,
the values of the various physical quantities are the same as the
upstream and downstream values for the J shock. Hence, on the large
scale the smaller substructure is unresolved, and the model appears as a
discontinuous MHD shock, such as that presented in Figure 9. 

This model demonstrates our ability to follow the formation
of multifluid MHD shocks, including short-lived
phenomena. Detailed studies, for a wide variety of
parameters and physical conditions, will be presented in
later work.


\section{Summary}

In this paper --- the first of a series --- we have discussed the 
formulation of a time-dependent, numerical MHD code designed to
model the formation and evolution of perpendicular MHD shocks
in dense molecular gas. Chief among the various features of our code is
that it is a multifluid code, incorporating the dynamics of neutral,
ion, and six different types of grain fluids. The grain fluids consist
of populations with two distinct radii, each with three different charge
states. In addition to including for the effects of elastic collisions
with gas particles, we also account for the transfer of mass and
momentum between the grain species due to capture of charged particles
on grain surfaces. 
We have presented the equations required to model this
physical system, and also described simplifying approximations
adopted within the numerical code.

To establish the accuracy of our numerical code, we have presented
several test models and compared their results to exact, analytical
solutions. In general, we found good agreement between the
numerical and analytical solutions. This included a test of the
formation of a C shock in a dust-free, weakly ionized gas, whose
analytical solution was determined by Chernoff (1987).

We also presented an illustrative model of a multifluid flow in which
inertia of the large grains was shown to be important over a finite
scale of the flow.
The possible effects that differential motion
of fluids in shocks may have on the state of the neutral gas and other
species will be investigated in more detail elsewhere. The effect of
grains (including their inertia) on the speed and propagation of signals
in dense gas will also be a topic of discussion in future work.
Using this numerical code as a tool to aid in further research, we
intend in subsequent papers to study the relevant physics and
chemistry of the formation and evolution of various types of nonlinear
MHD waves and shocks in molecular clouds, protostellar cores, and
disks.

\acknowledgements{We gratefully acknowledge support from the New York
Center for Studies on the Origins of Life (NSCORT), and the Dept. of
Physics, Applied Physics, and Astronomy at RPI, under NASA grant
NAG5-7598. Thoughtful comments by T. W. Hartquist are also gratefully
acknowledged.}


\clearpage
\appendix
\section{Chemical Network}
\samepage
\begin{deluxetable}{ll}
\tablehead{ \colhead{Reaction} & \colhead{Rate Coefficient (cm$^3$ s$^{-1}$)} }
\startdata
cosmic ray \plus  \Htwo    \goesto  \mplus    \plus  e         & a            \\
\mplus     \plus  e        \goesto  products                   & 
                                      $1.1\times 10^{-7}\,(300/T)$            \\
\mplus     \plus  a        \goesto  \aplus    \plus  m         & 
                                      $2.8\times 10^{-9}$                     \\
\mplus     \plus  \gminus  \goesto  m         \plus  \gzero    & b            \\
\mplus     \plus  \gzero   \goesto  m         \plus  \gplus    & b            \\
\mplus     \plus  \sgminus \goesto  m         \plus  \sgzero   & b            \\
\mplus     \plus  \sgzero  \goesto  m         \plus  \sgplus   & b            \\
\aplus     \plus  e        \goesto  a         \plus  \photon   & 
                                      $2.45\times 10^{-12}\,(300/T)^{0.86}$   \\
\aplus     \plus  \gminus  \goesto  a         \plus  \gzero    & b            \\
\aplus     \plus  \gzero   \goesto  a         \plus  \gplus    & b            \\
\aplus     \plus  \sgminus \goesto  a         \plus  \sgzero   & b            \\
\aplus     \plus  \sgzero  \goesto  a         \plus  \sgplus   & b            \\
e          \plus  \gzero   \goesto  \gminus                    & b            \\
\aplus     \plus  \gminus  \goesto  a         \plus  \gzero    & b            \\
\mplus     \plus  \gminus  \goesto  m         \plus  \gzero    & b            \\
e          \plus  \gplus   \goesto  \gzero                     & b            \\
e          \plus  \sgzero  \goesto  \sgminus                   & b            \\
e          \plus  \sgplus  \goesto  \sgzero                    & b            \\
\gminus    \plus  \gplus   \goesto  2\gzero                    & b            \\
\sgminus   \plus  \sgplus  \goesto  2\sgzero                   & b            \\
\enddata
\tbnt{a}{Ionization rate $\zeta$ with $\zeta\sim 10^{-17}$\,s$^{-1}$
    an adjustable parameter.}
\tbnt{b}{Rate coefficient computed by integrating cross sections from
    Draine \& Sutin (1987) over velocity distribution including relative
    drift.}
\end{deluxetable}



\begin{figure}
\figurenum{1a}
\plotone{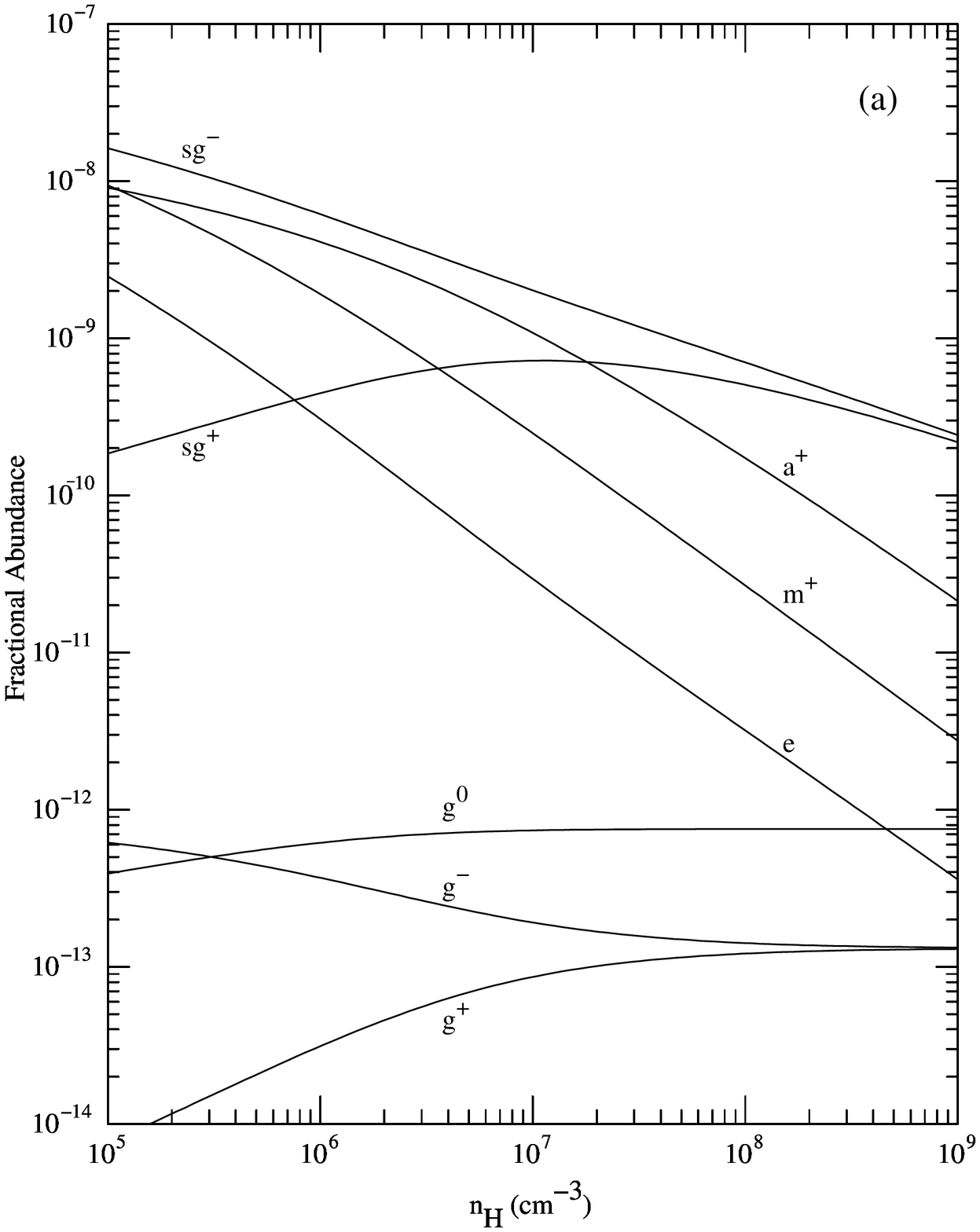}
\caption{Density-dependent fractional abundance of electrons (e),
atomic ions (a$^+$), molecular ions (m$^+$), small grains (sg) with radius
$\asg=4$\,\AA, and large grains (g) with radius $\ag=0.1$\,\micron,
as predicted by the simplified chemical network in Appendix A.
The total abundances of large- and small grains in all charge states
are $\xg=2\times 10^{-12}$ and $\xsg=2\times 10^{-7}$, respectively.}
\end{figure}

\begin{figure}
\figurenum{1b}
\plotone{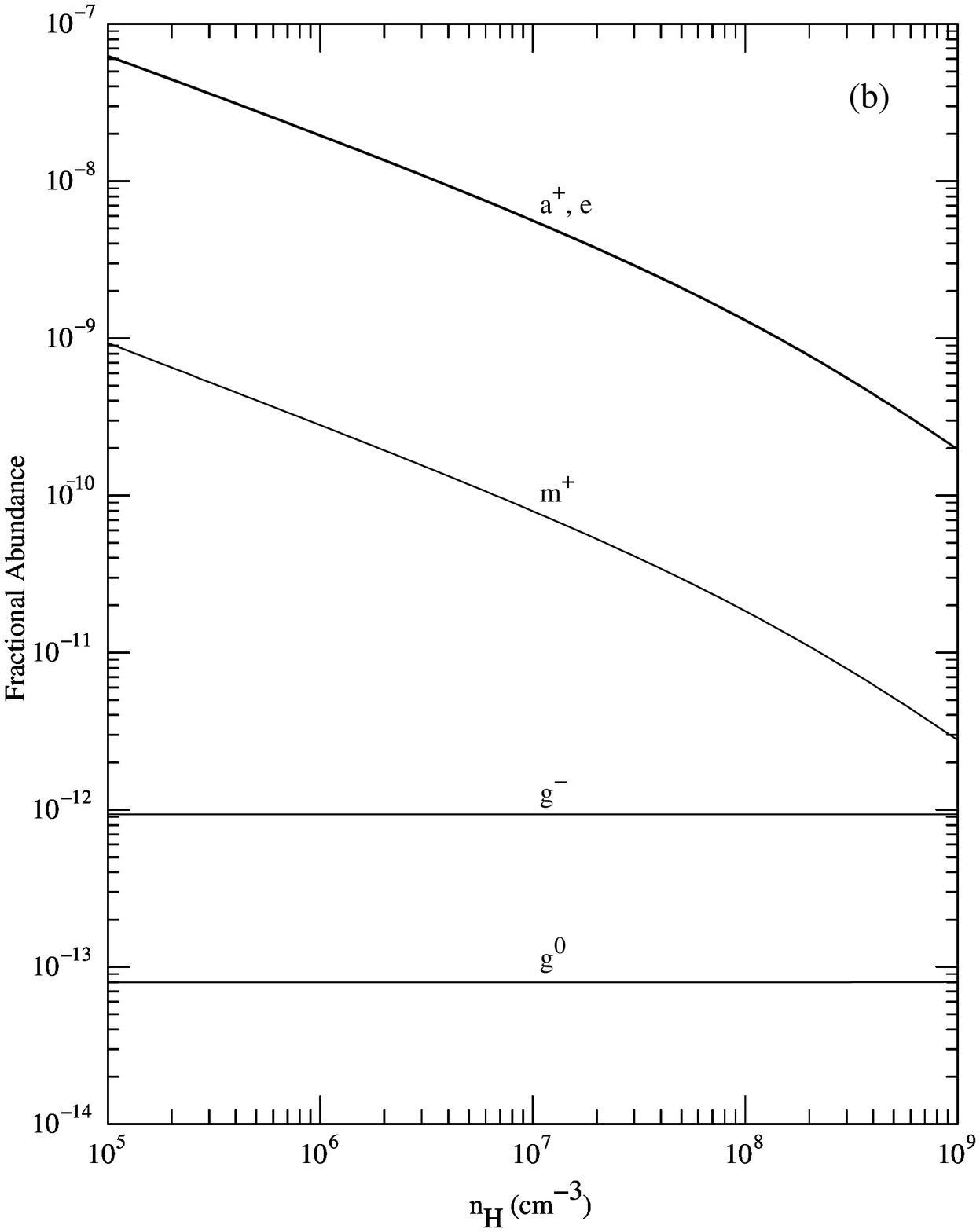}
\caption{Same as (a) but for $\xsg=0$.}
\end{figure}


\begin{figure}
\figurenum{2a}
\plotone{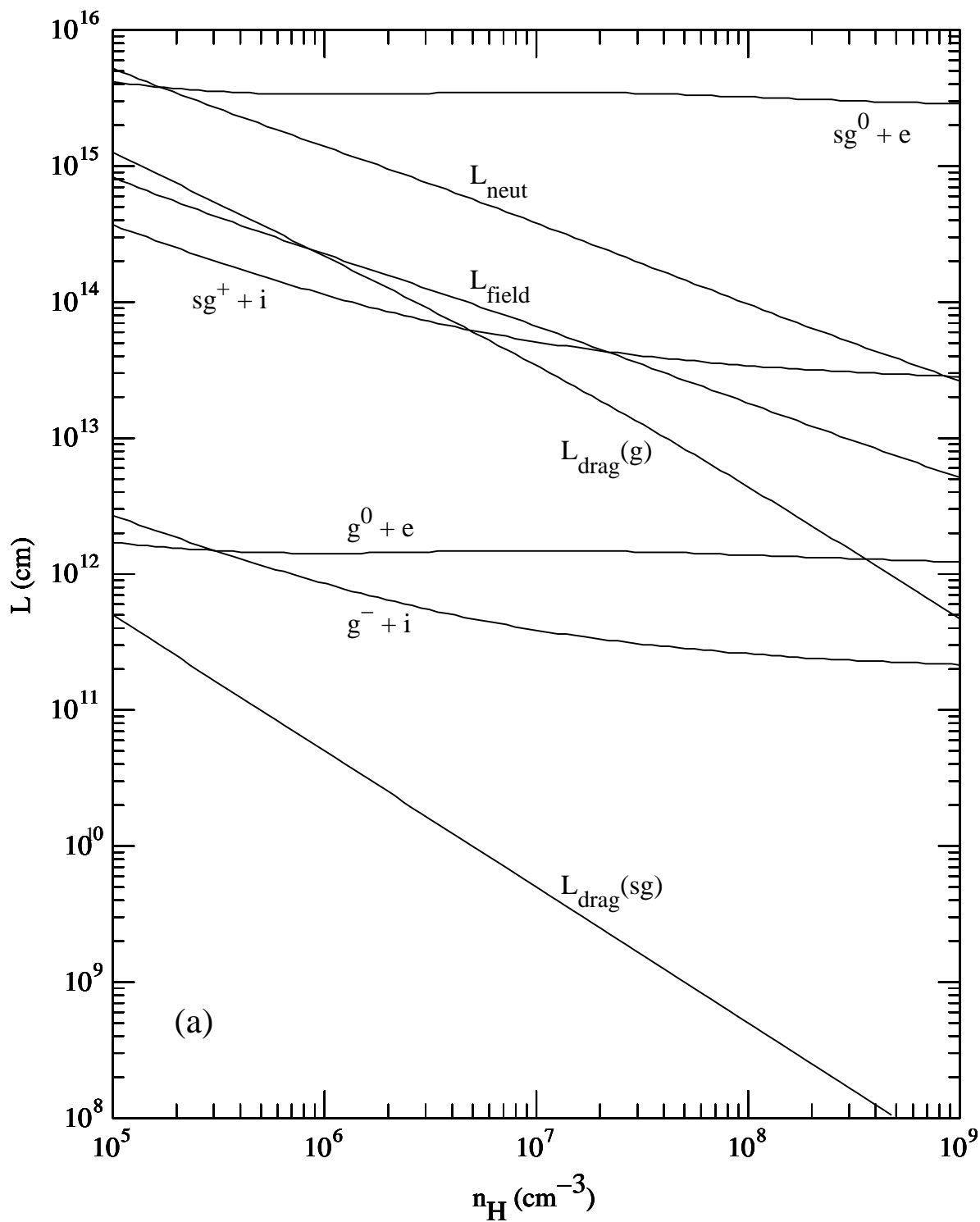}
\caption{Characteristic length scales for a steady shock with
speed $\vs=10~\kms$, plotted as a function of preshock density
(see text). Small grains are present with $\xsg=2\times 10^{-7}$.}
\end{figure}

\begin{figure}
\figurenum{2b}
\plotone{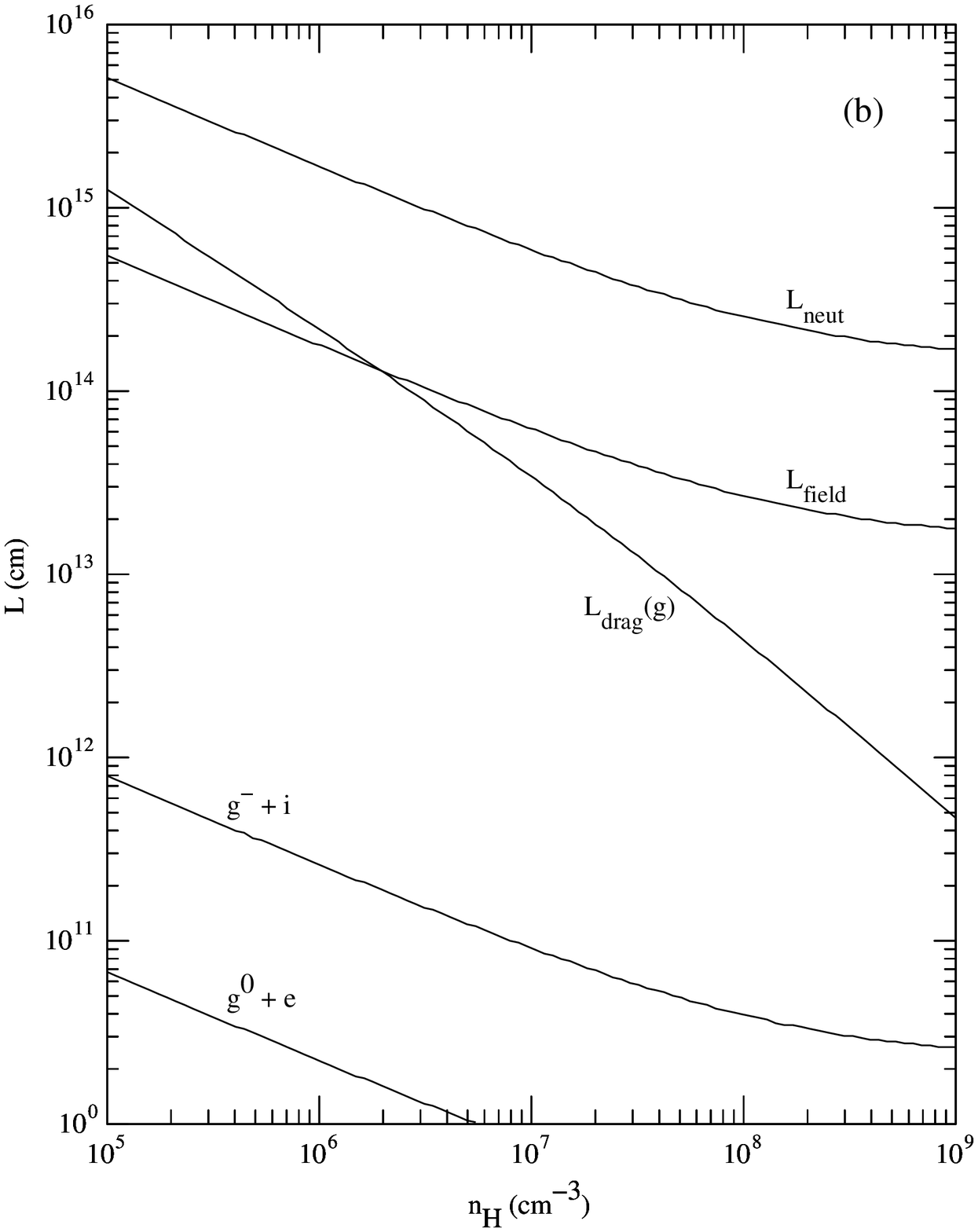}
\caption{Same as (a) but for $\xsg=0$.}
\end{figure}


\begin{figure}
\figurenum{3a}
\plotone{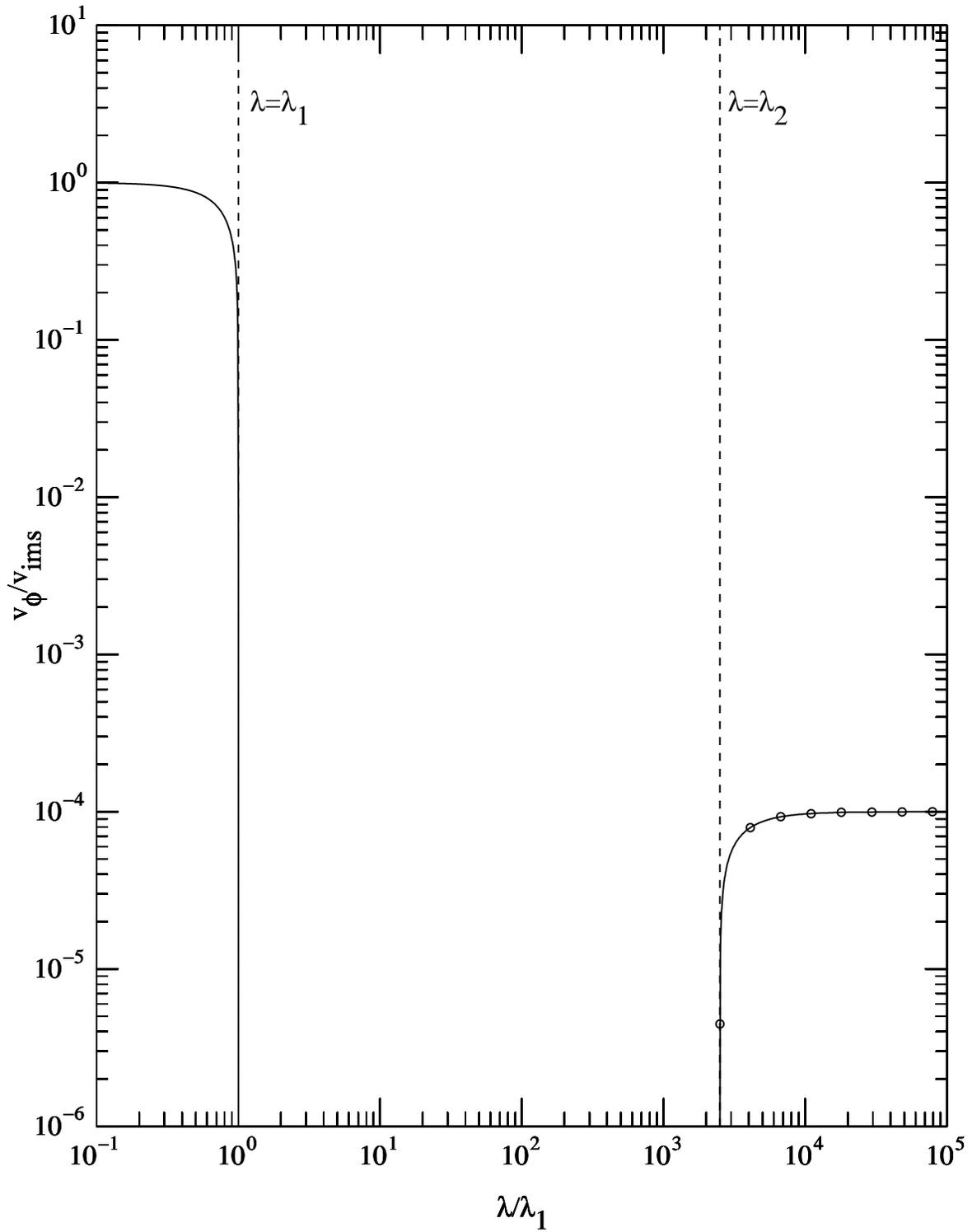}
\caption{Phase velocity of a small-amplitude wave propagating
perpendicular to the magnetic field in a cold ion+neutral plasma.
Solid curves: exact dispersion relation.
Open circles: effect of neglecting ion inertia.
}
\end{figure}


\begin{figure}
\figurenum{3b}
\plotone{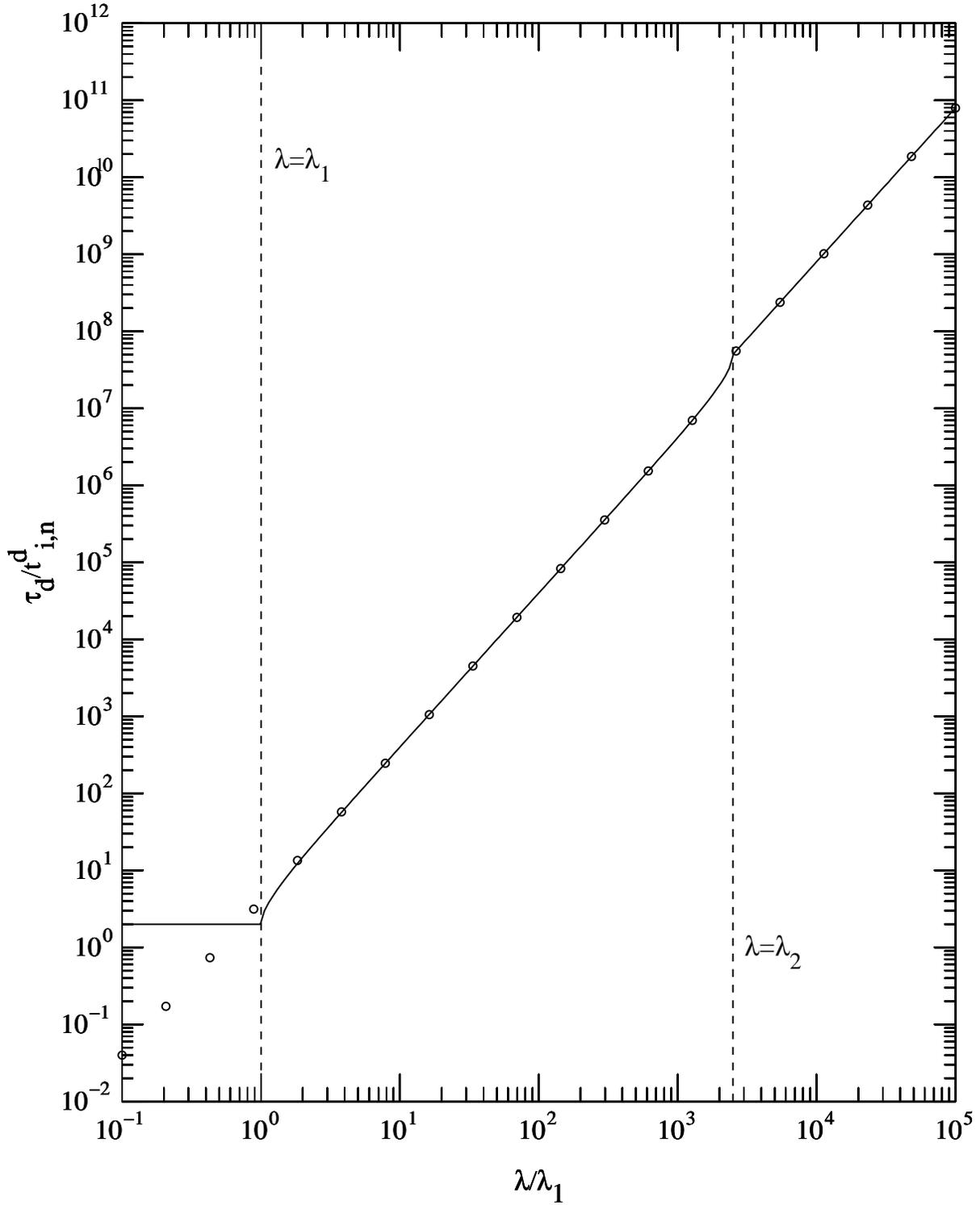}
\caption{Damping time, in units of the ion drag time,
for a small amplitude wave propagating
perpendicular to the magnetic field in a cold ion+neutral plasma.
Solid curves: exact dispersion relation.
Open circles: effect of neglecting ion inertia.
}
\end{figure}

\begin{figure}
\figurenum{4}
\epsscale{0.85}
\plotone{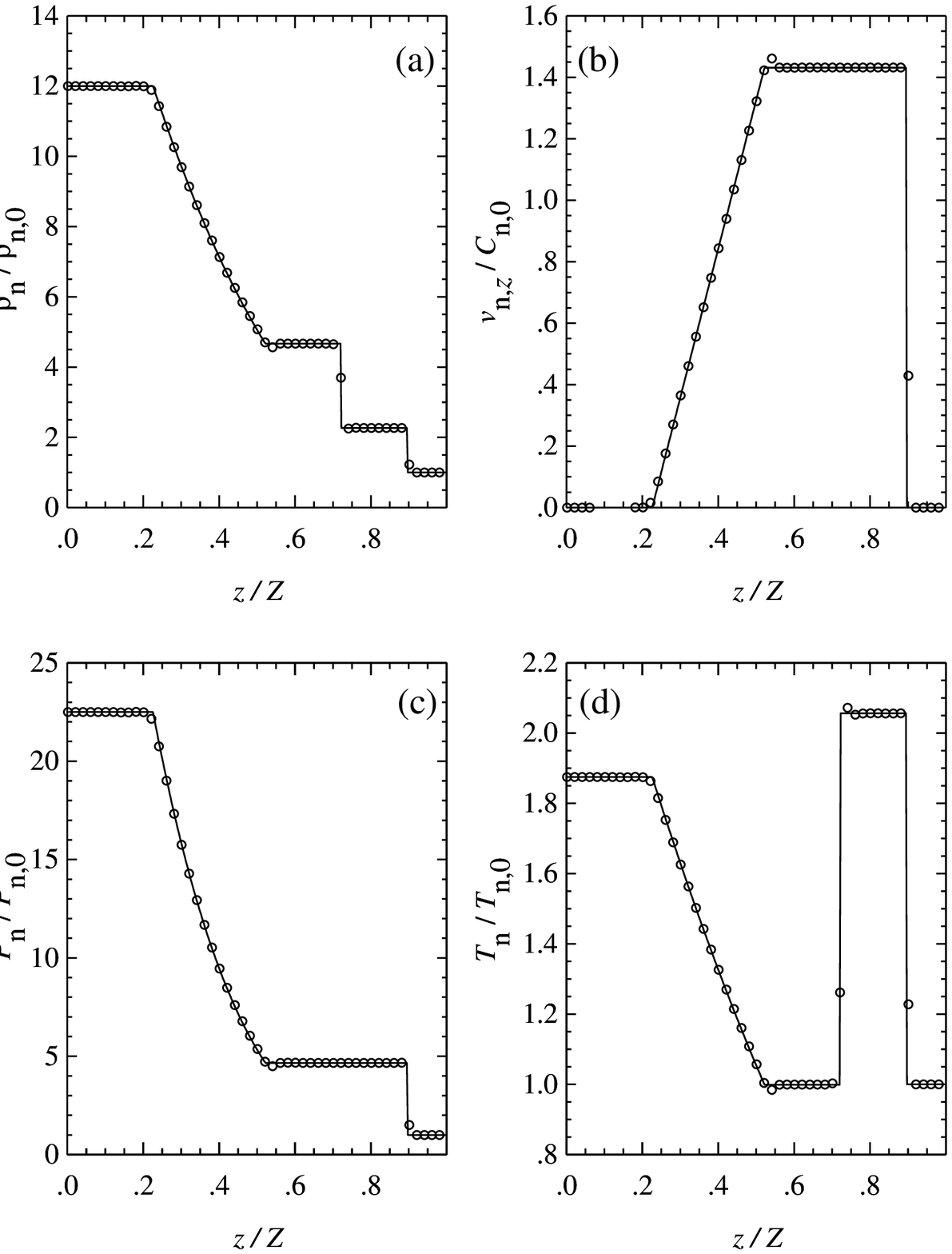}
\caption{Nonmagnetic shock tube problem at time $t_{1} > 0$.
Initially the fluid is stationary, with $\rhon/\rhono=12$,
$\Tn/\Tno = 1.25$ for $z/Z < 0.5$, and
$\rhon/\rhono=1$, $\Tn/\Tno = 1$ for $z/Z > 0.5$.
Circles: output from the numerical code described in this
paper. Solid curves: exact solution.
({\it a}) Density, as a function of position $z/Z$.
({\it b}) Gas velocity, in units of the isothermal speed of sound
$\Czero$ [$=(\kB \Tno/\mun)^{1/2}$] of the undisturbed gas.
({\it c}) Pressure. ({\it d}) Temperature. 
Since magnetic effects are not present in this particular model,
all of the fluids move together as a single fluid because of effective
collisional coupling. Note: for visual clarity, the curves showing the
numerical results use only one out of every eight data points.}
\end{figure}


\begin{figure}
\figurenum{5}
\epsscale{0.85}
\plotone{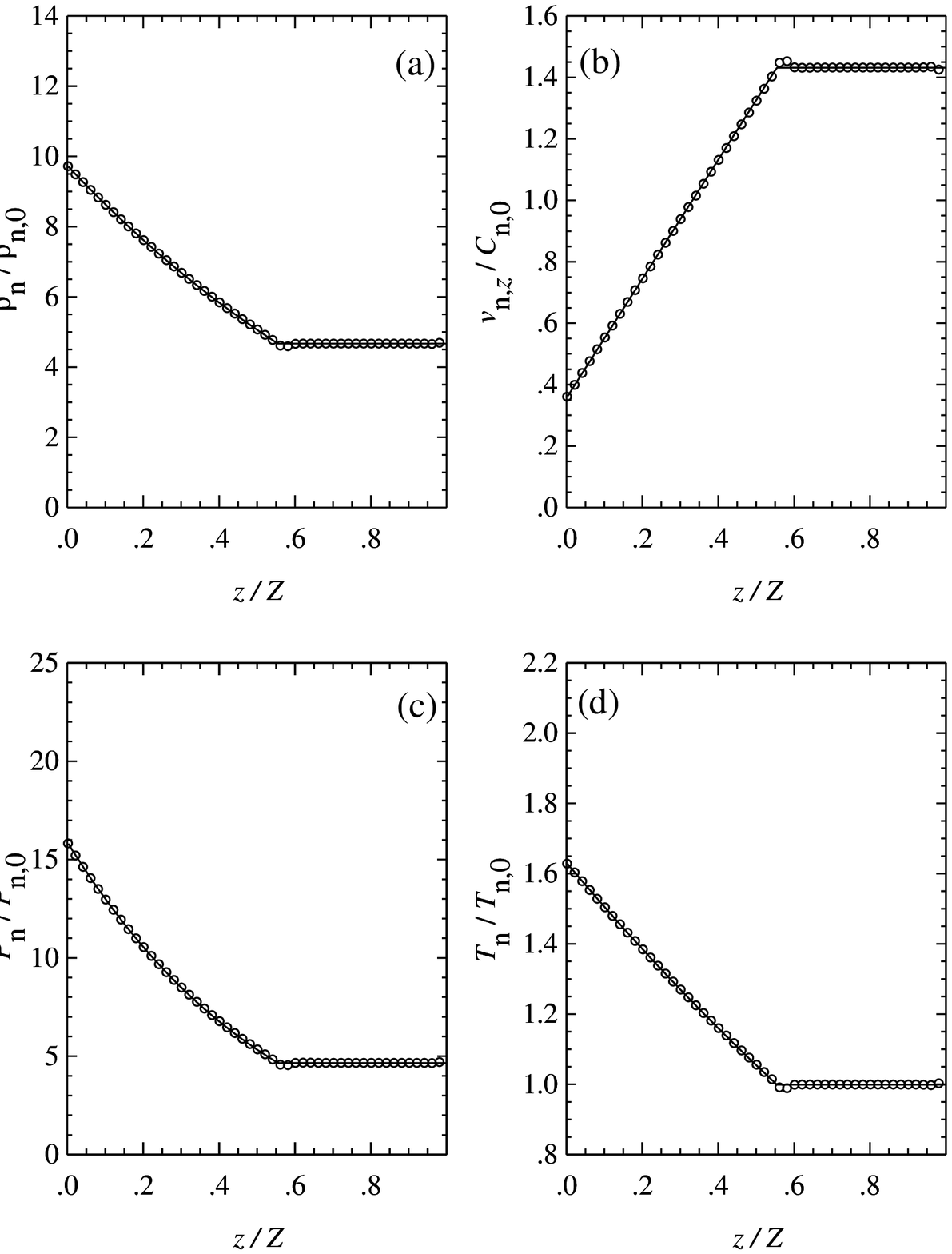}
\caption{Same as in Figure 4, except at a later time
$t_{2} > t_{1}$. By this time the forward-traveling shock wave and
contact discontinuity have passed beyond $z/Z = 1$, and the
backward-traveling rarefaction wave has extended to $z/Z < 0$. The
absence
of spurious numerical oscillations at $z/Z=0$ and $1$ is due to the
non-reflecting condition implemented at the boundaries.}
\end{figure}


\begin{figure}
\figurenum{6}
\epsscale{0.85}
\plotone{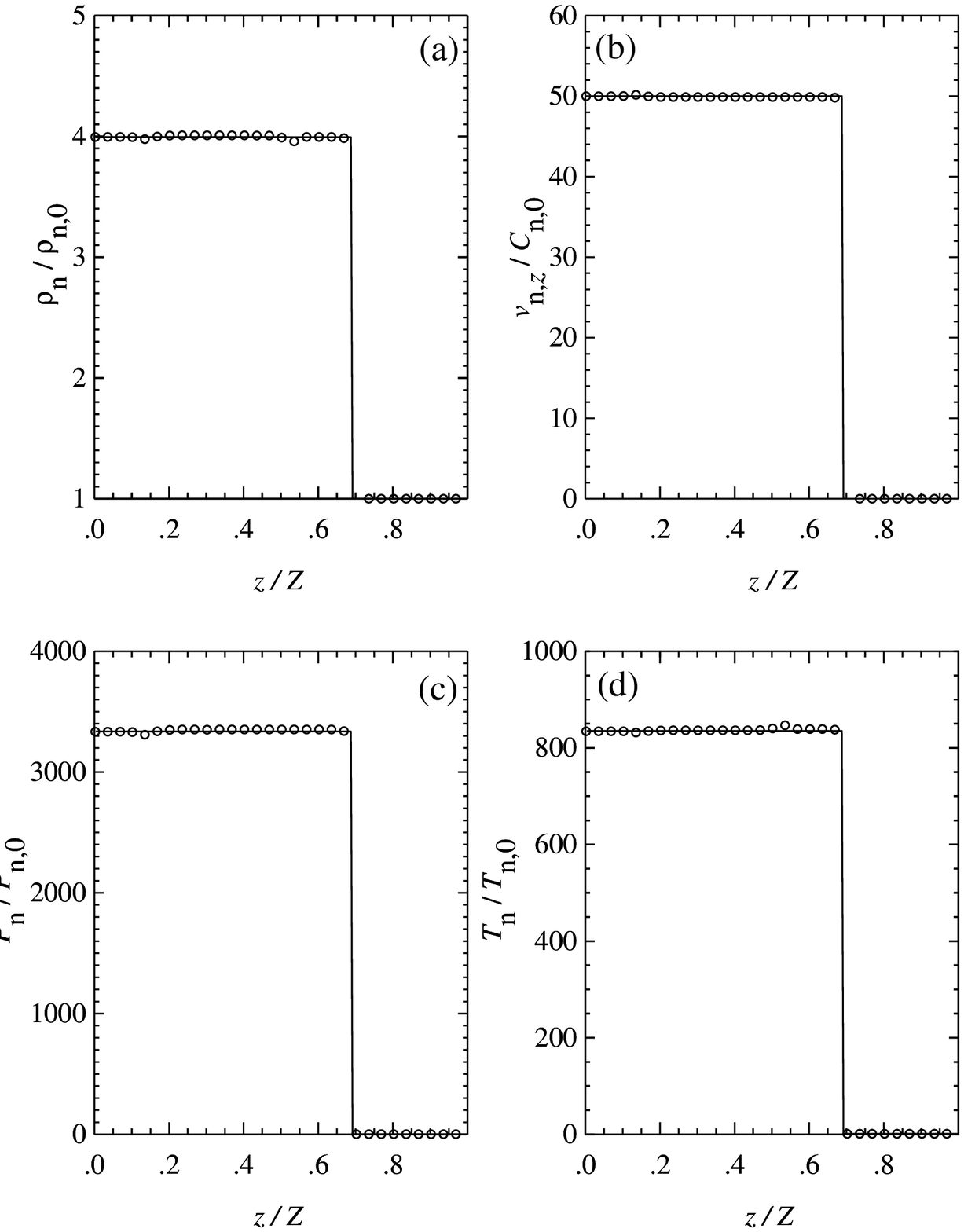}
\caption{High velocity, adiabatic, nonmagnetic inflow test. At time
$t=0$ a shock with velocity $\vnz = 50 \Czero$, $\rhon = 3.99 \rhono$,
and $\Tn = 835 \Tno$ enters the computational domain at $z=0$.
Shown is the model numerical solution (circles) at $t>0$, by which
time the shock has propagated into the computational domain.
({\it a}) Density, as a function of $z/Z$. ({\it b}) Velocity.
({\it c}) Pressure. ({\it d}) Temperature. The solid curves are the
exact solution, which are shown for comparison.}
\end{figure}


\begin{figure}
\figurenum{7}
\epsscale{0.85}
\plotone{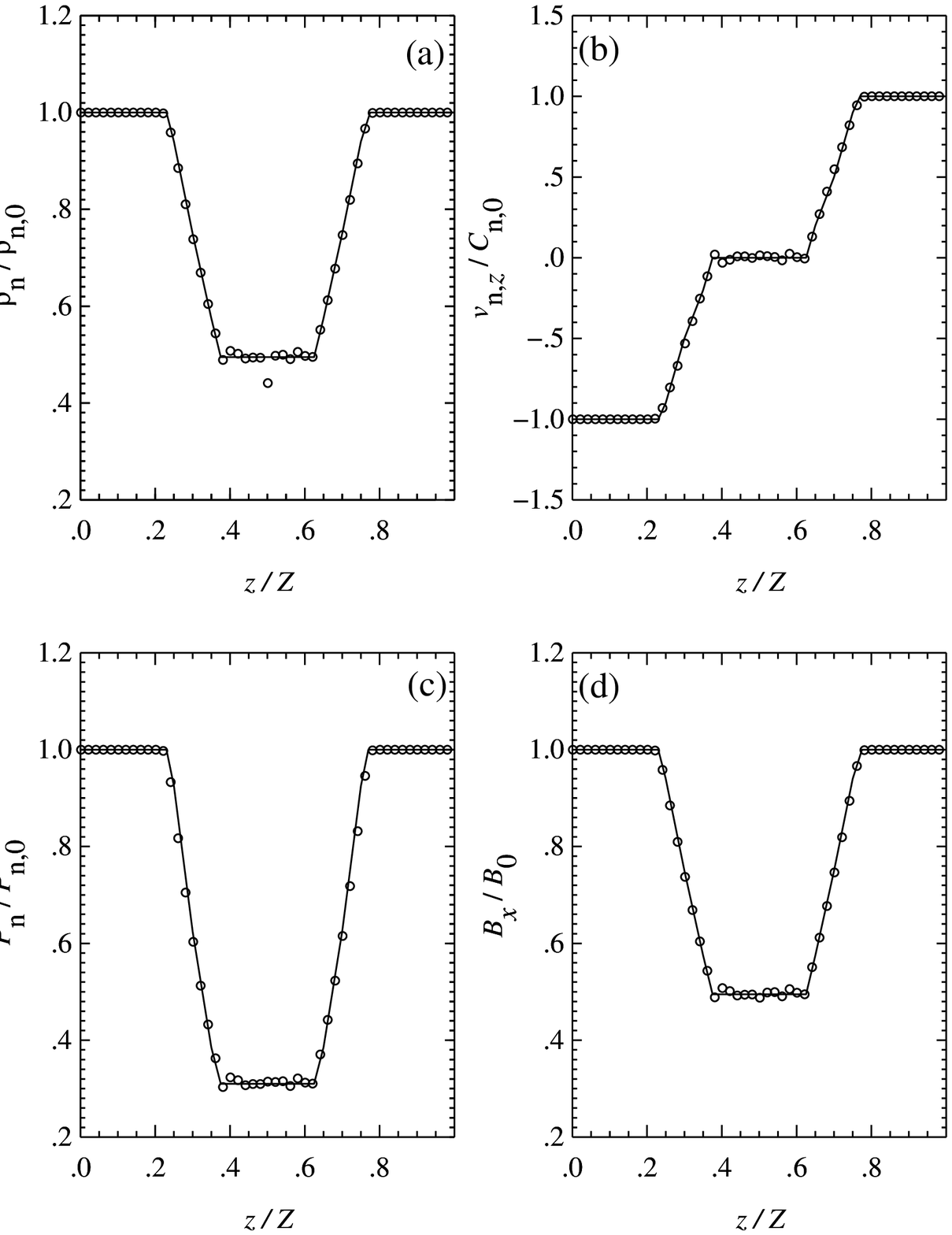}
\caption{Counterpropagating MHD rarefaction waves in a flux-frozen
fluid. Initially, $\rhon/\rhono = \Tn/\Tno = \Bx/\Bo = 1$,
$\vnz/\Czero = -1$ for $z/Z < 1/2$, and $\vnz/\Czero = 1$ for
$z/Z > 1/2$. Shown is the flow at a time $t > 0$, when the oppositely
traveling rarefaction waves have evacuated a significant amount of mass
and magnetic flux from the central
portion of the physical region. Circles represent the
numerical solution, which are compared with the MHD Riemann
solution (solid curves) of Ryu \& Jones (1995). Shown here are,
normalized to the initial state variables, the profiles of the
({\it a}) density, ({\it b}) velocity, ({\it c}) pressure, and
({\it d}) magnetic field at this time.}
\end{figure}


\begin{figure}
\figurenum{8}
\epsscale{0.80}
\plotone{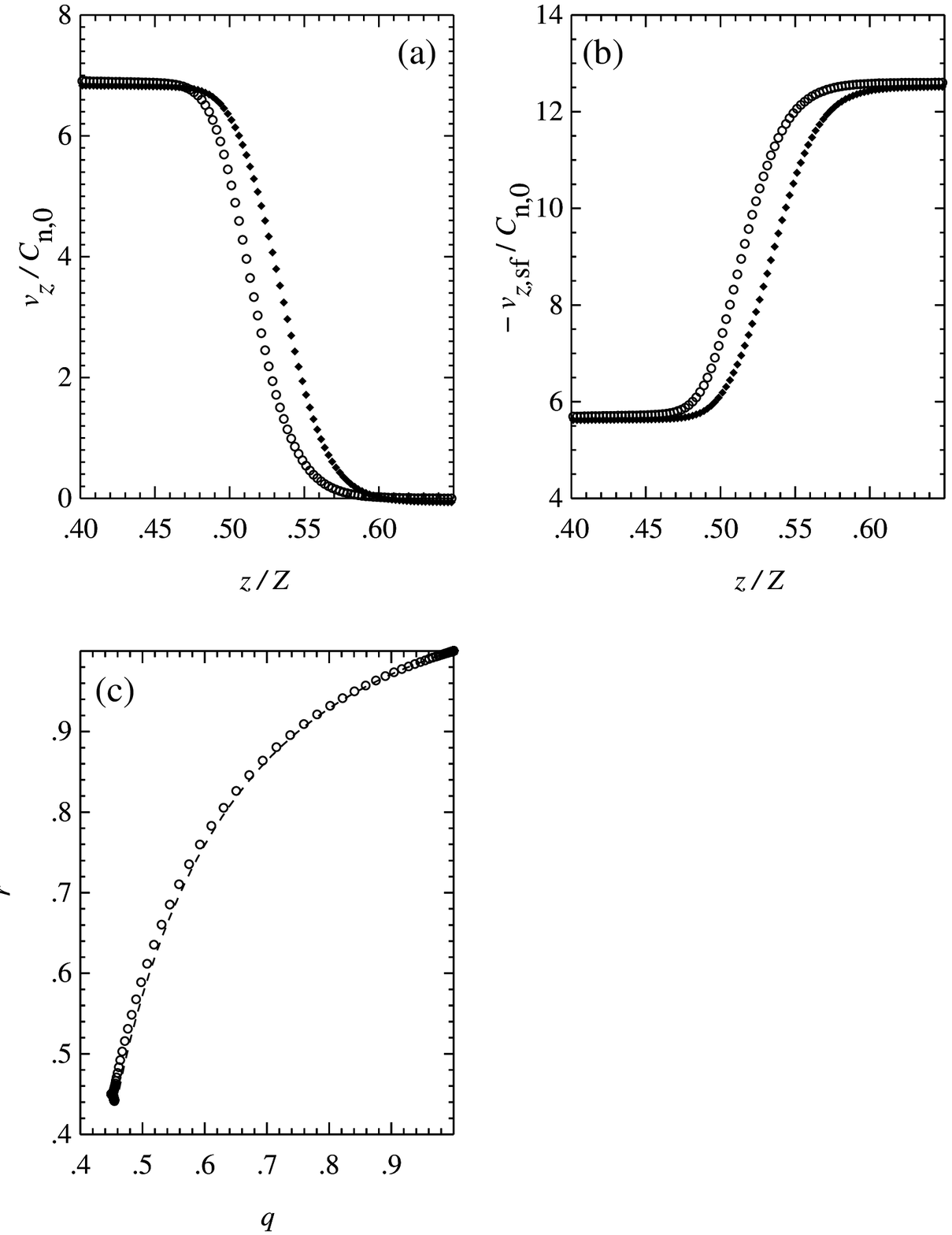}
\caption{An adiabatic MHD C shock in a weakly ionized gas,
created by an inflow J shock with $\machno = 9.75$ incident on the
boundary $z=0$ at time $t=0$. Model is shown after a steady C shock
is attained, with speed $\vshk = 12.6 \Czero$ and an Alfv\'{e}n Mach
number $\Mna = 2.04$.  ({\it a}) Velocities of the neutrals
(circles) and the ions (filled diamonds).
Characteristic of a C shock, there is a smooth, continuous transition
from supersonic to subsonic flow speeds in the neutral fluid. A magnetic
precursor in the ions runs ahead of the neutrals. ({\it b}) Velocities
of the neutrals $\vnzsf$ (circles) and ions $\vizsf$ (filled diamonds)
in the frame
comoving with the shock. Shown are the negatives of the relative
velocities $\vnzsf=(v_{z} - \vshk)$ and $\vizsf=(\viz - \vshk)$, 
at the same time $t$ as in ({\it a}). ({\it c}) Phase-plane variables
$r \equiv |\vnzsf|/\vshk$ and $q \equiv |\vizsf|/\vshk$. Circles:
numerical model results presented in ({\it b}). Dashed line: the
theoretical prediction for a hydromagnetic C shock with
$\Mna = 2.04$, calculated from equation (35) of Chernoff (1987).}
\end{figure}


\begin{figure}
\figurenum{9}
\epsscale{0.85}
\plotone{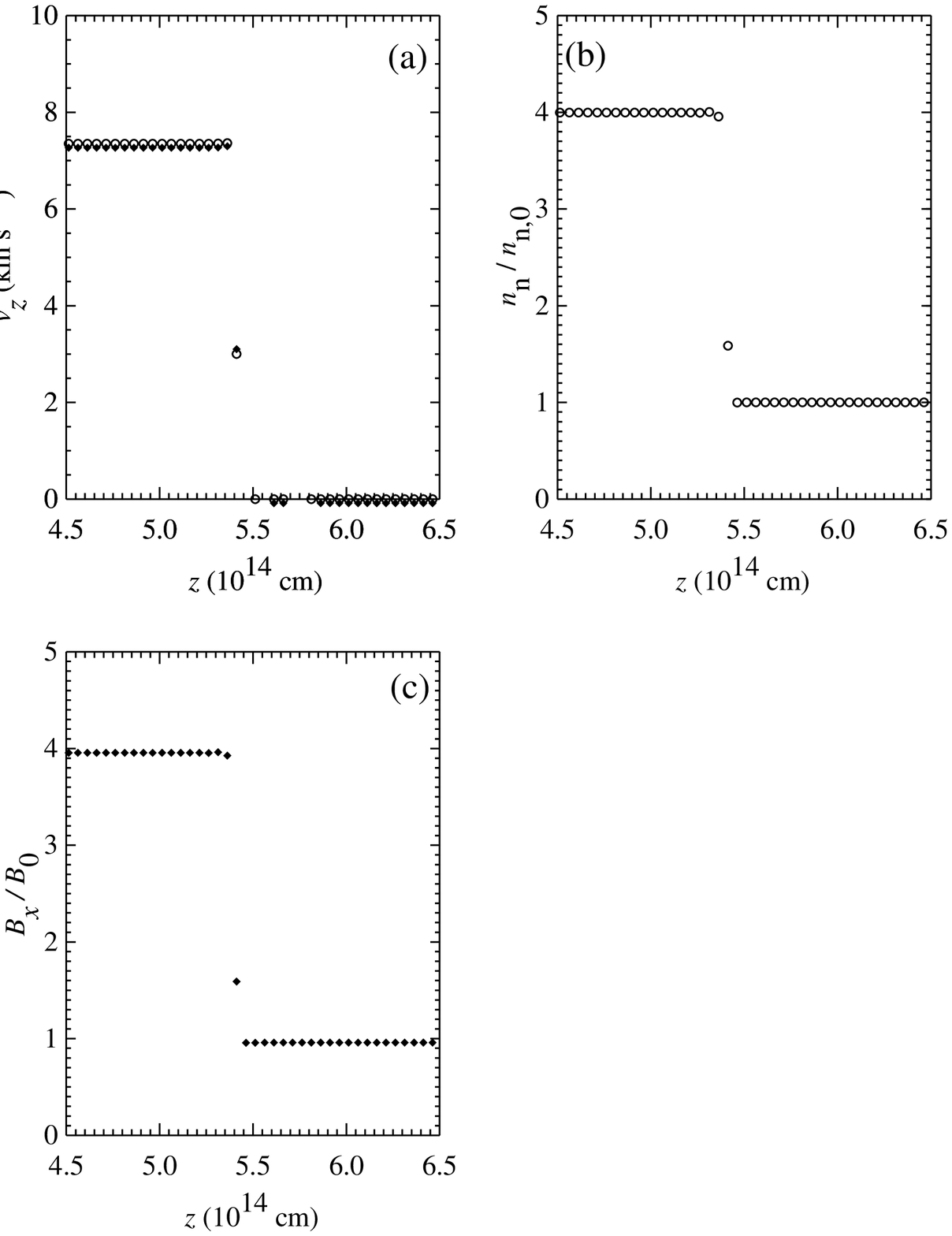}
\caption{Illustrative model with $\machno=40$, $\Mna = 115$.
The model is shown at a time $t=5.5 \times 10^8$ s.
({\it a}) Velocity of the neutrals (circles), and ions and small charged
grains (filled-triangles). On these scales the velocities of the large
grains are also the same as the neutrals, as are the small neutral grains.
({\it b}) Density of the neutrals, normalized to the upstream value
$\nno = 1.1 \times 10^{8}~\cc$. ({\it c}) Magnetic field strength,
normalized to $\Bo = 600~\mu{\rm G}$. On these length scales, all of
the fluids propagate as a J shock.} 
\end{figure}


\begin{figure}
\figurenum{10}
\epsscale{0.85}
\plotone{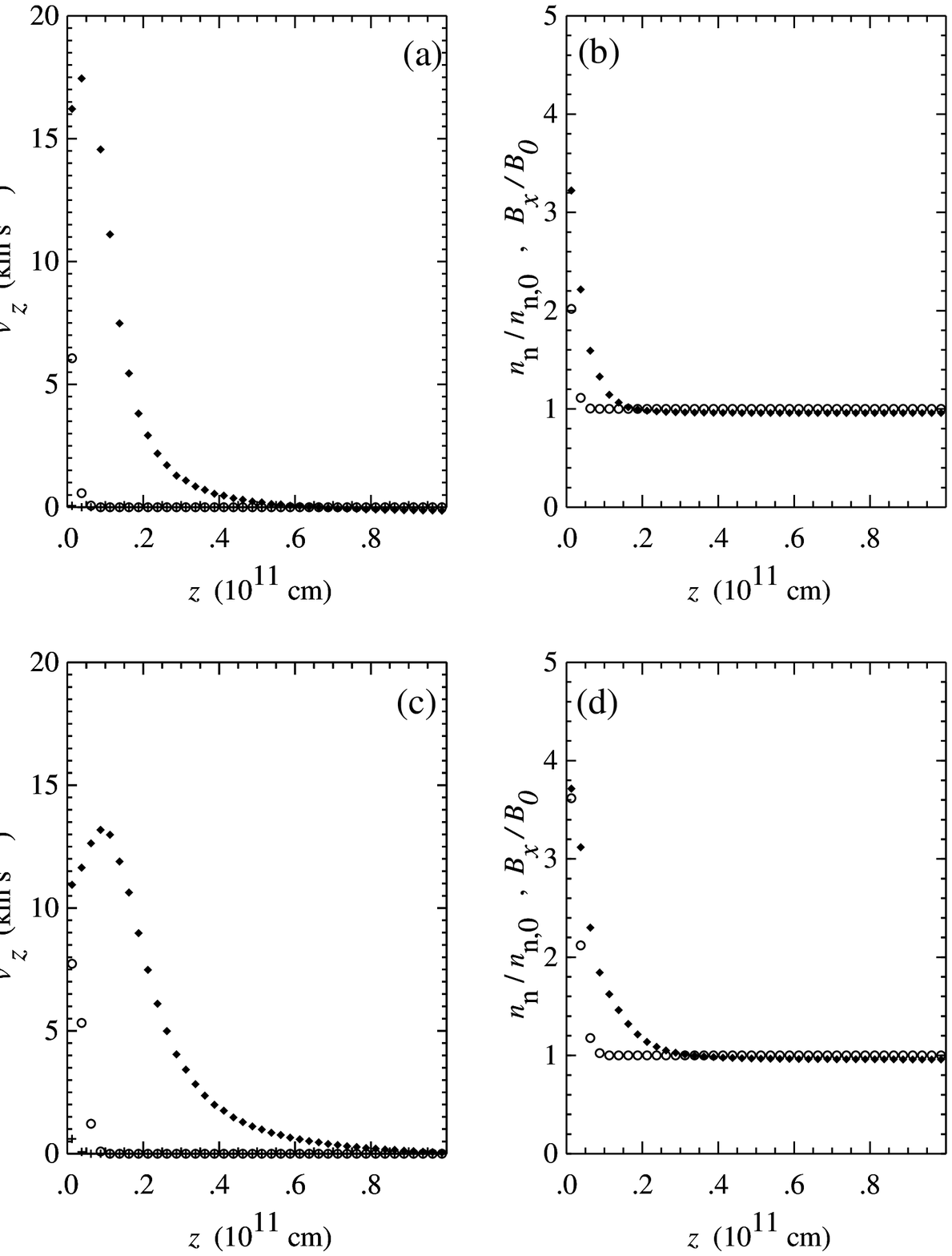}
\caption{Illustrative model, on much smaller length scales. The
parameters for this model are the same as in the preceding figure.
({\it a}) Velocities of the neutrals (circles), ions and small
charged grains (filled diamonds), and large dust grains (crosses) at
time $t=1.32 \times 10^3$ s. ({\it b}) Profile of the density
(circles, normalized to $\nno$) and magnetic field (filled diamonds,
in units of $\Bo$) at the same time as in ({\it a}). ({\it c}) Same as
in ({\it a}), except at $t=3.69 \times 10^3$ s. ({\it d}) Same as
({\it b}), but at the same time as ({\it c}). At these early times, the
magnetic field and the attached plasma species (ions, electrons, and
small charged grains) rapidly diffuse ahead of the imposed
flux-frozen flow for $z<0$ and $t<0$, at a rate much faster than the
inflowing neutrals and entrained large grains.}
\end{figure}


\begin{figure}
\figurenum{11}
\epsscale{0.85}
\plotone{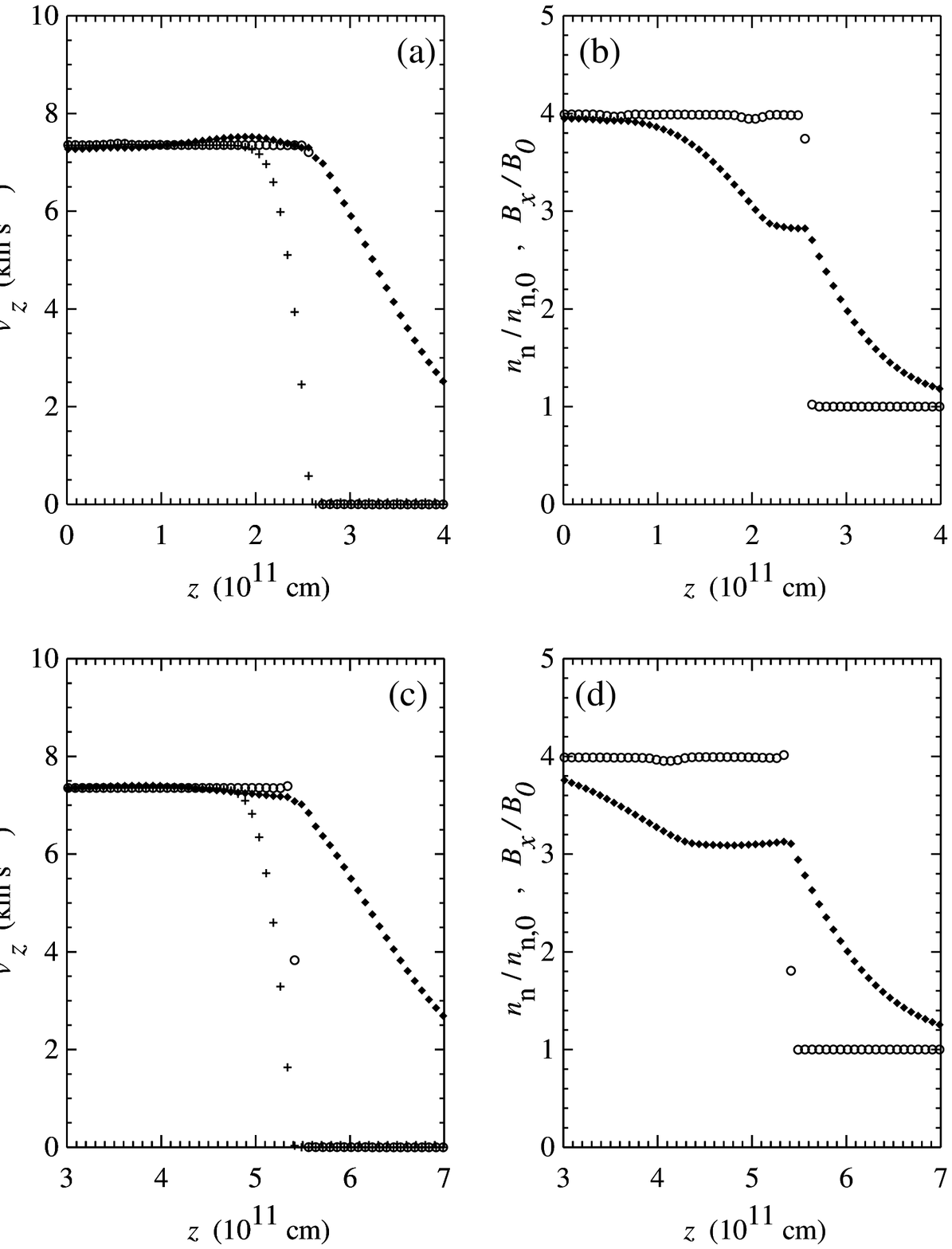}
\caption{Ilustrative model shock, same as Figure 10 (including symbols),
except at later times. ({\it a}) Velocities at $t=2.63 \times 10^5$s.
({\it b}) Density and magnetic field structure, at same time as in
({\it a}).
({\it c}) Velocities at $t=5.50 \times 10^5$s. ({\it d}) Density and
magnetic field strength, same time as in ({\it c}).}
\end{figure}


\end{document}